\documentclass[lettersize,journal]{IEEEtran}
\usepackage{amsmath,amsfonts}
\usepackage{amsthm}
\usepackage{booktabs}
\usepackage{array}
\usepackage{longtable}
\usepackage{amssymb}
\usepackage{textcomp}
\usepackage{stfloats}
\usepackage{url}
\usepackage{cite}
\usepackage{verbatim}
\usepackage{mathrsfs} 
\usepackage{graphicx}
\usepackage{xcolor}
\def\BibTeX{{\rm B\kern-.05em{\sc i\kern-.025em b}\kern-.08em
    T\kern-.1667em\lower.7ex\hbox{E}\kern-.125emX}}
\usepackage{balance}
\usepackage[font=footnotesize]{caption}
\usepackage{subcaption}
\usepackage{algorithm}
\usepackage{algpseudocode} 

\graphicspath{{ReachabilityTubes/}}

\usepackage{marginnote}

\newcommand{\real}{\mathbb{R}}

\newcommand{\bI}{\mathbf{I}}

\newcommand{\diag}{\mathrm{diag}}

\newcommand{\cF}{\mathcal{F}}
\newcommand{\cN}{\mathcal{N}}
\newcommand{\cG}{\mathcal{G}}
\newcommand{\cI}{\mathcal{I}}
\newcommand{\cL}{\mathcal{L}}
\newcommand{\cB}{\mathcal{B}}

\newcommand{\fs}{\textsf{s}}

\DeclareSymbolFont{bbold}{U}{bbold}{m}{n}
\DeclareSymbolFontAlphabet{\mathbbold}{bbold}

\DeclareMathAlphabet{\mymathbb}{U}{BOONDOX-ds}{m}{n}
\newcommand{\one}{\mymathbb{1}}
\newcommand{\zero}{\mymathbb{0}}

\newtheorem{theorem}{Theorem}
\newtheorem{lemma}{Lemma}

\newtheorem{problem}{Problem}

\allowdisplaybreaks[3]

\title{Reachability Analysis for  Power Systems with Heterogeneous Resources via Jordan Transformation}

\author{Damola Ajeyemi, Antonin Colot, Sairaj Dhople, Emiliano Dall'Anese, and Saber Jafarpour \thanks{D. Ajeyemi and E. Dall'Anese are with the Division of Systems Engineering, Boston University. Antonin Colot is with Elia Grid International. Sairaj Dhople is with the Department of Electrical and Computer Engineering, University of Minnesota. Saber Jafarpour is with the Department of Computer Science, University of Colorado Boulder. Corresponding author: D. Ajeyemi, email: dajeyemi@bu.edu.}}

\begin{document}

\maketitle
\begin{abstract}
This paper develops a computationally efficient framework for reachability analysis of transmission-level power system dynamics with synchronous generators, grid-forming and grid-following inverters, and uncertain power injections/withdrawals. Starting from reduced-order device models and a frequency-divider representation, we derive a linear ordinary-differential-equation model suitable for efficient reachable-set computation under bounded disturbances across network buses. The proposed reachability method combines interval reachability and contraction-based bounds to construct certified over-approximations for the linear ordinary-differential-equation model. A real Jordan transformation separates non-oscillatory modes, handled through a linear embedding system, from oscillatory modes, enclosed using contraction-based ball bounds. Numerical experiments on a modified IEEE 39-bus system validate the reachable tubes against high-fidelity  electromagnetic-transient (EMT)  simulations, and demonstrate multi-second reachable sets computed in sub-second time.

\end{abstract}

\section{Introduction} \label{sec:intro}
Power systems are experiencing increasing variability due to the rapid growth of inverter-based resources and large-scale electrified loads such as data centers~\cite{TAYLOR20161322,milano2018foundations,dorfler2023control}. System operators can benefit significantly from tools that, in near real time, reliably estimate the range of possible operating conditions and transient responses following combinations of disturbances and events~\cite{Dhople-SetTheoretic-2016,Lara-2024}. Traditional tools for small-signal analysis characterize local modal properties around an operating point, but are  not designed to certify behavior under large disturbances. Conversely, transient stability analysis relies on time-domain simulation, which produces individual trajectories but cannot characterize worst-case behavior over uncertainty sets~\cite{kundur2007power,sauer2017power}. As modern grids experience increasing variability, operators must reason about sets of possible frequencies, powers, and flows  emerging from variations of loads across the system, rather than individual simulated scenarios.

Reachability analysis provides a model-based framework for bounding the evolution of uncertain dynamical systems~\cite{althoff2021set,chen2022reachability}. Instead of simulating a single trajectory, sets of states are propagated through the dynamics to obtain a \emph{reachable tube} containing all trajectories induced by admissible bounded disturbances. In power systems, these tubes can estimate, e.g., worst-case frequency deviations, rate of change of frequency~(RoCoF), and settling time following disturbances; such elements are key to certifying performance and ensuring secure operation. Because exact reachable-set computation for power transmission systems is generally intractable, existing methods either consider short horizons or employ computationally expensive set representations~\cite{ADG-2012,althoff2014formal,pico2013reachability,Choi-CDC-2015,Choi-2017,el2017compositional,jin2010reachability,zhang2020review}. The former may miss the full post-disturbance evolution from transient response to settling, while the latter can scale poorly with system dimension. Zonotope-based methods, differential-inclusion methods, and Hamilton--Jacobi techniques~\cite{bansal2017hamilton} can therefore be computationally prohibitive for large networks.

This paper develops a computationally efficient reachability framework for transmission-level power systems with heterogeneous devices, including synchronous generators and grid-forming and grid-following inverters. Starting from a differential--algebraic network representation, we derive a reduced linear dynamic model using a frequency-divider representation and Schur-complement reduction. Reachable sets are then computed through a hybrid interval--contraction method based on a real Jordan decomposition of the dynamics. Non-oscillatory modes are bounded using interval analysis, while oscillatory modes are enclosed through contraction-based ball bounds. The resulting reachable tube contains all trajectories of the reduced model induced by admissible bounded disturbances. High-fidelity EMT simulations are used to assess how accurately these reduced-model tubes capture the behavior of a detailed dynamic model. Thus, the proposed method provides certified over-approximations for the analytical model, while the EMT simulations validate the modeling approximation and the practical usefulness of the computed tubes.

The main contributions are summarized as follows:

\noindent \emph{c1)} To address the high dimensionality and differential--algebraic structure of power-system models, we derive a tractable analytical model using a frequency-divider representation and eliminate the algebraic network variables to obtain a reduced linear model. The resulting formulation captures the dominant electromechanical and inverter dynamics, together with the network power-flow coupling, while remaining tractable for reachability analysis.

\noindent \emph{c2)} Based on the derived linear model, we develop a mode-decoupled reachability framework that combines interval bounds for non-oscillatory modes with contraction-based bounds for oscillatory modes. For piecewise-constant post-event inputs, we further introduce a steady-state-informed construction based on an equilibrium shift and shifted homogeneous dynamics, and intersect the resulting tube with the transient tube to reduce conservatism. The method enables fast computation of reachable sets over long time intervals.

\noindent \emph{c3)} We validate the reduced-model reachable tubes against detailed dq0 electromagnetic-transient (EMT) simulations~\cite{lara2023powersimulationsdynamics} that capture heterogeneous device dynamics of synchronous generators and grid-forming and grid-following inverters. The comparison assesses whether the certified tubes computed for the analytical model provide accurate and practically useful envelopes for trajectories of the high-fidelity EMT model.

The case studies show that the proposed framework provides accurate bounds on system trajectories while maintaining significantly lower computational cost than existing methods. In particular, the framework computes multi-second reachable tubes in approximately $0.1~\mathrm{s}$ for the modified IEEE 39-bus system, which is several orders of magnitude faster than high-fidelity EMT simulation.

\section{Problem Statement} \label{sec:prob_statement}
We consider a power transmission system with buses $\mathcal{N}=\{1,\ldots,N\}$ and  transmission lines $\mathcal{E}\subseteq \mathcal{N}\times\mathcal{N}$.\footnote{\emph{Notation}. The set of real numbers, positive real numbers, and nonnegative real numbers are denoted by $\real$, $\real_{>0}$, and $\real_{\geq 0}$, respectively. Similarly, we define $\mathbb{Z}$, $\mathbb{Z}_{>0}$, and $\mathbb{Z}_{\geq 0}$ as the sets of integers, positive integers, and nonnegative integers. For $n \in \mathbb{Z}_{>0}$, $\zero_n$ is the $n$-dimensional zero vector and $\one_n$ a vector of all ones; similarly, for $n, m \in \mathbb{Z}_{>0}$, $\zero_{n \times m}$ is a matrix of dimensions $n \times m$ with all zeros. The $n \times n$ identity matrix is denoted by $\bI_n$, and $e_i \in \real^n$ represents the $i$th canonical basis vector. For a given vector $x \in \real^n$ and matrix $X \in \real^{n \times m}$, $x^\top$ and $X^\top$ denote transposition; if $X \in \real^{n \times n}$ is invertible, $X^{-1}$ denotes its inverse. The diagonal and block-diagonal constructions are written as $\diag(a_1, \ldots, a_n)$ for scalars $a_i \in \real$ and $\operatorname{blkdiag}(X_1, \ldots, X_k)$ for compatible matrices $X_i$. The trace, determinant, and rank of a matrix $X$ are denoted by $\operatorname{tr}(X)$, $\det(X)$, and $\operatorname{rank}(X)$. The spectrum of a square matrix $A$ is denoted by $\operatorname{spec}(A)$, and its spectral radius by $\rho(A) = \max_{\lambda \in \operatorname{spec}(A)} |\lambda|$. For a matrix $M$ and set $\mathcal S$, we define $M\mathcal S:=\{Mx:\;x\in\mathcal S\}$. For $\lambda \in \mathbb{C}$, $\Re(\lambda)$ and $\Im(\lambda)$ denote the real and imaginary parts of $\lambda$, respectively. A real matrix $A \in \real^{n \times n}$ is said to be Hurwitz if $\max\{\Re(\lambda) : \lambda \in \operatorname{spec}(A)\} < 0$. For a finite set $\cN \subset \mathbb{Z}$, $|\cN|$ denotes its cardinality. Given $x \in \mathbb{R}^n$, $x\in[\underline{x},\overline{x}]$ denotes the hyper-rectangle $[\underline{x},\overline{x}] := \{\,x\in\mathbb{R}^n \mid \underline{x}_i \le x_i \le \overline{x}_i,\; i=1,\ldots,n\}$. For a vector or matrix $M$, $[M]^+$ and $[M]^-$ denote its elementwise positive and negative parts. Also, $\mathcal{B}(r,c):=\{y\in\real^q:\|y-c\|_2\le r\}$ denotes the closed Euclidean ball of radius $r$ centered at $c$.} The system features heterogeneous resources, including synchronous generators (SGs) and inverter-based resources (IBRs). The overall model is described by a differential–algebraic system 
\begin{align}
\dot x_\fs=f(x_\fs,y_\fs,u_\fs),\qquad 
\zero=g(x_\fs,y_\fs,u_\fs),  \label{eq:DAE}
\end{align}
where $x_\fs$ collects dynamic states of lines and devices, $y_\fs$ collects algebraic variables (bus voltages and currents), and $u_\fs$ represents inputs such as power references, loads, and non-controllable generation. The algebraic equations arise from Kirchhoff’s current law at each bus and from load models.

Each SG contributes electromechanical and control dynamics, including the rotor angle and speed governed by swing equations, together with turbine-governor, excitation-system, and stabilizer dynamics~\cite{kundur2007power,ajala2020library}. Models for IBRs capture both the power-electronic interface and internal controls. An averaged voltage-source-converter~(VSC) representation typically models the conversion of a DC-link voltage to an AC terminal voltage through an $LC$ or $LCL$ filter whose currents and voltages form dynamic states. Control strategies include:~(i) \emph{grid-following~(GFL)} control, which synchronizes through a phase-locked loop~(PLL) and tracks active and reactive power references; and~(ii) \emph{grid-forming (GFM)} control, which regulates an AC voltage waveform and can provide frequency and voltage support through, e.g., droop, virtual inertia, and virtual oscillator-based strategies~\cite{Venkat-2022,dorfler2023control}. 

During system operation, it is important to characterize the evolution of selected differential and algebraic states under disturbances (e.g., load variations and generation fluctuations) and discrete events (e.g., generator or load trips). These uncertainties are typically modeled via an admissible input set $\mathcal{U}$. In this paper, we provide a solution to the following problem:

\begin{problem}[\emph{Power System Evolution}]
\label{prob:evolution}
Given a set of admissible inputs $\mathcal{U}$ describing disturbances or events in the power system~\eqref{eq:DAE}, develop methods that can efficiently compute tight estimates of the evolution of the dynamic states $x_\fs$ and the algebraic states $y_\fs$ in real-time.
\end{problem}

Problem~\ref{prob:evolution} is naturally addressed through the notion of \emph{reachable sets} for the power system dynamic model~\eqref{eq:DAE}. Intuitively, the reachable set characterizes all states that the system can attain under admissible disturbances and events; in practice, one often focuses on selected states of interest, such as angles, frequencies, and active-power variables. Formally, the reachable set of system~\eqref{eq:DAE} at time $t$ is the set of all dynamic states, $x_\fs(t)$, and all algebraic states, $y_\fs(t)$, that satisfy power dynamics~\eqref{eq:DAE} for some admissible disturbance, $u_\fs:[t_0,t]\to\mathcal U$.  Computing the exact reachable set of the power system~\eqref{eq:DAE} is generally intractable. Therefore, our objective is to construct tight over approximations in a computationally efficient manner that enable real-time analysis of the system's dynamic behaviors.

\section{Models}
\label{sec:models}

Due to the high dimensionality and DAE structure of the EMT dynamic power system model~\eqref{eq:DAE}, reachability analysis is computationally prohibitive. Therefore, the methodology proposed in this paper follows a two-pronged modeling approach: 

\noindent $\triangleright$ A lower-order analytical model capturing dominant electromechanical, inverter, and power-flow dynamics is used for algorithmic design due to its computational tractability.

\noindent $\triangleright$ The proposed reachability method is validated through high-fidelity EMT simulations that include detailed device and network dynamics~\cite{lara2023powersimulationsdynamics}. 

\begin{table}[b!]
\centering
\renewcommand{\arraystretch}{1.06}
\setlength{\tabcolsep}{5pt}
\normalsize
\resizebox{\columnwidth}{!}{%
\begin{tabular}{l | l | l}
\hline
\textbf{Subsystem} & \textbf{High-Fidelity dq0 EMT Model} & \textbf{Analytical Model} \\
\hline
Generator
  & Flux + Shaft + AVR + Governor + PSS
  & Swing equation + Governor lag \\
\hline
Transmission Line
  & Dynamic Branch $\pi$-model
  & DC approximation \\
\hline
Load
  & ZIP model (Constant-power Loads)
  & ZIP model (Constant-power Loads) \\
\hline
GFL IBR
  & \begin{tabular}[t]{@{}l@{}}PLL + PI-based $P/Q$ outer control\\+ current control inner loop + LCL filter\end{tabular}
  & Algebraic model via frequency divider \\
\hline
GFM IBR
  & \begin{tabular}[t]{@{}l@{}}Virtual inertia and $Q$-droop outer control\\+ voltage control inner loop + LCL filter\end{tabular}
  & Virtual inertia swing equation \\
\hline
\end{tabular}%
}
\caption{dq0 EMT models used for validation (explained in Section~\ref{sec:simulations}) and analytical model  used for the algorithmic design (explained in Section~\ref{sec:models}).}
\label{tab:emt_vs_analytical}
\end{table}

Table~\ref{tab:emt_vs_analytical} summarizes the correspondence between the high-fidelity EMT models used for validation and the simplified analytical models used for algorithmic design. In the following subsections, we describe the simplified lower-order analytical model used to develop the reachability method.

\subsection{Dynamic and algebraic models for resources}
Let $\mathcal{G}\subset\mathcal{N}$ denote buses hosting synchronous generators, $\mathcal{I}\subset\mathcal{N}$ buses hosting grid-forming IBRs, $\mathcal{F}\subset\mathcal{N}$ buses hosting grid-following IBRs, and $\mathcal{L}\subset\mathcal{N}$ load buses. For simplicity of notation, we assume that $\mathcal G,\mathcal I,\mathcal F,\mathcal L$
form a partition of $\mathcal N$. The model can be extended to cases where a bus hosts multiple assets, at the expense of much heavier notation. Let $\theta_{\text{bus}} \in \mathbb{R}^{N}$ denote the vector of bus-voltage phase angles. The transmission network $(\mathcal{N},\mathcal{E})$ is connected and $\mathcal{G}\cup\mathcal{I}\neq\emptyset$; i.e., there is at least one generator or one GFM IBR. Without loss of generality, bus $N$ is chosen as the reference bus, and its angle is fixed: $\theta_N = 0$. Bus numbering is such that the angle reference bus is $N \in \mathcal{G}\cup\mathcal{I}$. The vector of independent bus-angle variables is therefore $\theta = [\theta_1,\ldots,\theta_{N-1}]^\top \in \mathbb{R}^{N-1}$. Define the embedding matrix $I_0 =
\begin{bmatrix}
I_{N-1}\\
\zero_{1\times(N-1)}
\end{bmatrix}
\in\mathbb{R}^{N\times(N-1)}$, so that the full vector of bus angles can be written as $\theta_{\text{bus}} = I_0\,\theta$. Let $\omega = [\omega_1,\ldots,\omega_N]^\top \in \mathbb{R}^{N}$ collect the bus angular frequency deviations from the nominal synchronous electrical radian frequency, $\omega_\mathrm{s}$.

Transmission lines are assumed to be predominantly inductive. 
For each bus $i\in\mathcal{N}$, let $\mathcal{N}_i=\{j\in\mathcal{N}:(i,j)\in\mathcal{E}\}$ denote the set of neighboring buses. Under the AC power-flow model, the active-power injection at bus $i$ depends nonlinearly on voltage magnitudes and angle differences across incident lines. Assuming small voltage-angle differences, constant voltage magnitudes, and negligible line resistances, the standard DC power-flow approximation yields 
\begin{equation*}
p_i(\theta)
=
\sum_{j\in\mathcal{N}_i} b_{ij}(\theta_i-\theta_j),    
\end{equation*}
where $b_{ij}=1/x_{ij}>0$ denotes the line susceptance associated with edge $(i,j)\in\mathcal{E}$. Stacking the bus power injections as $p(\theta)=[p_1(\theta),\ldots,p_N(\theta)]^\top$, the DC network model can be written compactly as $p(\theta) = \mathcal{B}\,\theta_{\text{bus}}$, where $\mathcal{B}\in\mathbb{R}^{N\times N}$ is the network susceptance (Laplacian) matrix~\cite{kundur2007power}. Using the angle reference-bus representation $\theta_{\text{bus}} = I_0\theta$, the DC network power injections satisfy $p(\theta) = \mathcal{B} I_0 \theta$. Next, we describe the model of each asset. Subscripts are dropped to ease the notational burden. 

\subsubsection{Grid-Forming IBRs} 
Each GFM IBR is modeled as a controlled voltage source characterized by an internal angle, $\theta$, and deviation from synchronous frequency, $\omega$. Let $p$ denote the active-power injection of the GFM IBR, and let $\hat p$ be a filtered measurement of $p$. The inverter regulates its frequency according to the frequency–power droop law
\begin{equation}
    \omega 
    = -\,m_{p}\big(\hat p - p^{\mathrm{ref}}\big),
    \label{eq:gfm_droop}
\end{equation}
where $p^\mathrm{ref}$ is the active-power reference and $m_{p}>0$ is the droop slope. Since the raw power measurement may contain oscillatory components and feature noise attributable to switching, the controller employs a low-pass filter $T_{f}\dot{\hat p} 
    = -\,\hat p + p$,
where $T_{f}>0$ is the measurement-filter time constant. Differentiating both sides of~\eqref{eq:gfm_droop} and substituting the filter dynamics yields
\begin{equation}
    M\dot{\omega} 
    = -\,D\omega 
      + p^\mathrm{ref} - p,
    \label{eq:gfm_swing_pre_balance}
\end{equation}
with positive constants $M$ and $D$ that are functions of control and filter parameters:  $M = \tfrac{T_{f}}{m_{p}}$ and $D = \tfrac{1}{m_{p}}$, which follows from a widely recognized equivalence relationship between virtual-synchronous-machine control and droop control~\cite{Suul-Equivalence-2014}.  If a static load $p_{\ell}$ is also connected at the bus and $p(\theta)$ denotes the active power exported from the bus to the rest of the network, the bus power balance enforces $p = p_{\ell} + p(\theta)$. Collecting the angle and frequency dynamics, we have
\begin{subequations}
\label{eq:dynamics_ibr}
\begin{align}
    \dot{\theta} &= \omega,
    \label{eq:dynamics_ibr_angle}\\
    M\dot{\omega} &= -\,D\omega 
    + p^\mathrm{ref} - p_{\ell} - p(\theta).
    \label{eq:dynamics_ibr_freq}
\end{align}
\end{subequations} 

\subsubsection{Grid-Following IBRs} 
For buses hosting GFL IBRs, the inverter synchronizes to the grid frequency through a PLL. The GFL modulates its active-power injection in proportion to the PLL-estimated frequency deviation to provide so-called fast frequency response. We model the GFL inverter as an algebraic power injection depending on the local frequency deviation. Specifically, the active-power reference is adjusted according to $p \approx p^\mathrm{ref} - K\omega$, where $p^\mathrm{ref}$ is the active-power setpoint and $K>0$ quantifies the frequency-sensitivity gain for fast frequency response. Active-power balance imposes $p = p_{\ell} + p(\theta)$, where $p_{\ell}$ denotes the local load and $p(\theta)$ denotes the active power exported to the network. With these constructs in place, the GFL model takes the algebraic form:
\begin{equation}
0
= p^\mathrm{ref} - K\,\omega
  - p_{\ell} - p(\theta).
\label{eq:gfl_balance}
\end{equation}

\subsubsection{Synchronous generators} 
For buses hosting synchronous generators, we adopt the standard reduced-order electromechanical model~\cite{ajala2020library}
\begin{subequations}
\label{eq:dynamics_generator}
\begin{align}
    \dot{\theta}&= \omega,\\
    M\,\dot{\omega} 
        &= -\,D\,\omega 
           + p_{\mathrm m} 
           - p_{\ell} 
           - p(\theta),\\
    \tau\,\dot{p}_{\mathrm m} 
        &= -\,p_{\mathrm m} 
           + p^\mathrm{ref} 
           - R\,\omega ,
\end{align}
\end{subequations}
where $\theta$ and $\omega$ denote the rotor angle and frequency deviation relative to the angle reference bus. The parameters $M>0$ and $D>0$ denote (with a slight abuse of notation) the generator inertia and damping coefficients, respectively. The variable $p_{\mathrm m}$ denotes the turbine mechanical power, and $p^\mathrm{ref}$ is the generator active-power reference (typically determined by secondary control or economic dispatch). The parameters $\tau>0$ and $R>0$ denote the turbine time constant and the inverse frequency–power droop coefficient, respectively. The electrical power exported to the network is denoted by $p(\theta)$. 

\subsubsection{Loads} 
Loads are modeled as constant active-power consumptions: $0 = -\,p_{\ell} - p(\theta)$, where $p_{\ell}$ denotes the net active-power demand at the bus. 

\subsection{System-level DAE model} 
To write the system model compactly, define the selector matrices
$S_\cG\!\in\!\{0,1\}^{N\times|\cG|}$,\;
$S_\cI\!\in\!\{0,1\}^{N\times|\cI|}$,\;
$S_\cF\!\in\!\{0,1\}^{N\times|\cF|}$,\;
$S_\cL\!\in\!\{0,1\}^{N\times|\cL|}$
that extract the corresponding bus variables from vectors defined on the space of all buses. For example, $\omega_\mathcal{G}=S_\mathcal{G}^\top\omega$, 
$\omega_\mathcal{I}=S_\mathcal{I}^\top\omega$,
$\omega_\mathcal{F}=S_\mathcal{F}^\top\omega$. Similarly, $S_{\mathcal{GI}} = [\,S_{\mathcal G}\ \ S_{\mathcal I}\,]\in\{0,1\}^{N\times(|\mathcal G|+|\mathcal I|)}$, $S_{\mathcal{LF}} = [\,S_{\mathcal F}\ \ S_{\mathcal L}\,]\in\{0,1\}^{N\times(|\mathcal F|+|\mathcal L|)}$.
The matrices 
$\bar S_\cG\in\{0,1\}^{(N-1)\times|\cG|}$,
$\bar S_\cI\in\{0,1\}^{(N-1)\times|\cI|}$,
$\bar S_\cF\in\{0,1\}^{(N-1)\times|\cF|}$, and
$\bar S_\cL\in\{0,1\}^{(N-1)\times|\cL|}$
denote the analogous selectors acting on the reduced
(non-angle-reference) angle vector $\theta\in\mathbb R^{N-1}$; e.g., $\theta_\cG=\bar S_\cG^\top\theta$. We define $\bar S_{\mathcal{GI}}=[\,\bar S_\cG\ \bar S_\cI\,]$ and $\bar S_{\mathcal{LF}}=[\,\bar S_\cF\ \bar S_\cL\,]$. Recall that $\theta_{\text{bus}}=I_0\theta$;  then, the DC active-power injections satisfy $p(\theta) = \mathcal{B}\theta_{\text{bus}} = \mathcal{B}I_0\theta = \bar{\mathcal{B}}\theta$, where $\bar{\mathcal{B}}=\mathcal{B}I_0$. The power injections associated with each bus class are therefore $p_\mathcal{G}(\theta)=S_\mathcal{G}^\top\bar{\mathcal{B}}\theta$, $p_\mathcal{I}(\theta)=S_\mathcal{I}^\top\bar{\mathcal{B}}\theta$, $p_\mathcal{F}(\theta)=S_\mathcal{F}^\top\bar{\mathcal{B}}\theta$, $p_\mathcal{L}(\theta)=S_\mathcal{L}^\top\bar{\mathcal{B}}\theta$.

One critical task is to relate the frequencies at the buses with GFL IBRs and with loads to the frequencies in buses with swing-type dynamics. To this end, we leverage the notion of frequency divider~\cite{milano2016frequency}. Since the PLL dynamics of GFL IBRs are significantly faster than the electromechanical time scales considered here, the GFL frequencies can be approximated as algebraic functions of the swing-bus frequencies. This relation can be derived from the quasi-stationary power-flow conditions at the non-swing buses. Let $\theta_{\mathcal{GI}} =
\begin{bmatrix}
\theta_{\mathcal{I}}^\top \;
\theta_{\mathcal{G}}^\top
\end{bmatrix}^{\!\top}$, $\theta_{\mathcal{LF}} =
\begin{bmatrix}
\theta_{\mathcal{F}}^\top \;
\theta_{\mathcal{L}}^\top
\end{bmatrix}^{\!\top}$  denote the stacked swing-bus and non-swing bus angles, respectively (with the reference angle removed). For future developments, it is convenient to define $C_{\mathcal F}=\big[\,I_{|\mathcal F|}\ \ \zero_{|\mathcal F|\times|\mathcal L|}\,\big]\in\mathbb R^{|\mathcal F|\times(|\mathcal F|+|\mathcal L|)}$, which denotes the selector that extracts the GFL components from a stacked non-swing vector $[\cdot_\mathcal F^\top\ \cdot_\mathcal L^\top]^\top$; finally, $I_{-1}$ is defined as $I_{-1}=
\begin{bmatrix}
I_{|\cG|+|\cI|-1} & -\,\one_{|\cG|+|\cI|-1}
\end{bmatrix}
\in \real^{(|\cG|+|\cI|-1)\times(|\cG|+|\cI|)}$. Then, the frequency–divider matrix
\begin{equation}
H
=
-\,C_{\mathcal{F}}
\left(S_{\mathcal{LF}}^\top\bar{\mathcal{B}}\bar S_{\mathcal{LF}}\right)^{-1}
S_{\mathcal{LF}}^\top\bar{\mathcal{B}}\bar S_{\mathcal{GI}}
I_{-1}
\label{eq:H_matrix}
\end{equation}
is well-defined and uniquely maps the swing-bus frequency deviations
to the GFL IBR bus frequencies $\omega_{\mathcal F}$ as: 
\begin{equation}
\label{eq:divider}
\omega_{\mathcal{F}} = H_{\mathcal{I}}\omega_{\mathcal{I}} + H_{\mathcal{G}}\omega_{\mathcal{G}}
\end{equation}
with $H=
\begin{bmatrix}
H_{\mathcal I} & H_{\mathcal G}
\end{bmatrix}
\in \mathbb{R}^{|\mathcal F|\times (|\mathcal G|+|\mathcal I|)}$. To establish this result, recall that the DC power-flow relation is \(p^{\mathrm{DC}}(\theta)=\bar{\mathcal{B}}\theta\), and that 
$\theta
=
\bar S_{\mathcal{GI}}\theta_{\mathcal{GI}}
+
\bar S_{\mathcal{LF}}\theta_{\mathcal{LF}}$. Then, it follows that $p_{\mathcal{LF}}^{\mathrm{DC}}(\theta)
=
S_{\mathcal{LF}}^\top\bar{\mathcal{B}}\bar S_{\mathcal{GI}}\theta_{\mathcal{GI}}
+
S_{\mathcal{LF}}^\top\bar{\mathcal{B}}\bar S_{\mathcal{LF}}\theta_{\mathcal{LF}}$, where 
\begin{equation*}
    p_{\mathcal{LF}}^{\mathrm{DC}}(\theta) = \begin{bmatrix}
p^\mathrm{ref}_{\mathcal F}-K_{\mathcal F}\omega_{\mathcal F}-p_{\ell,\mathcal F}\\
-\,p_{\ell,\mathcal L}
\end{bmatrix}
\end{equation*} 
stacks the power injections at the buses $\cF$ and $\cL$. Next, differentiate both sides with respect to time. Assuming that the PLL dynamics are much faster than generator frequency dynamics and using a quasi-steady-state approximation (i.e., GFL IBR frequencies behave like algebraic variables, not dynamic states), we have $\dot{p}_{\mathcal{LF}}^{\mathrm{DC}}(\theta) = 0$. Then, we obtain 
\begin{equation*}
    - S_{\mathcal{LF}}^\top\bar{\mathcal{B}}\bar S_{\mathcal{GI}} \dot{\theta}_{\mathcal{GI}} = 
S_{\mathcal{LF}}^\top\bar{\mathcal{B}}\bar S_{\mathcal{LF}}\dot{\theta}_{\mathcal{LF}}.
\end{equation*}
Under the topological assumptions, matrix $S_{\mathcal{LF}}^\top\bar{\mathcal{B}}\bar S_{\mathcal{LF}}$ is invertible (see Section~\ref{sec:InvertibilityForFD}). Therefore, we get 
\begin{equation*}
    \dot\theta_{\mathcal{LF}}
=
-
\left(S_{\mathcal{LF}}^\top\bar{\mathcal{B}}\bar S_{\mathcal{LF}}\right)^{-1}
S_{\mathcal{LF}}^\top\bar{\mathcal{B}}\bar S_{\mathcal{GI}}
\dot\theta_{\mathcal{GI}}.
\end{equation*}
Since
$\dot{\theta}_{\mathcal{LF}}=\omega_{\mathcal{LF}}$,
$\dot{\theta}_{\mathcal{GI}}=I_{-1}\omega_{\mathcal{GI}}$, and
$\omega_{\mathcal F}=C_{\mathcal F}\omega_{\mathcal{LF}}$, it follows that
$\omega_{\mathcal F}
=
-\,C_{\mathcal F}
\left(S_{\mathcal{LF}}^\top\bar{\mathcal B}\bar S_{\mathcal{LF}}\right)^{-1}
S_{\mathcal{LF}}^\top\bar{\mathcal B}\bar S_{\mathcal{GI}}
I_{-1}\omega_{\mathcal{GI}}$.
Using the definition of $H$ in~\eqref{eq:H_matrix}, together with
$H=[\,H_{\mathcal I}\ \ H_{\mathcal G}\,]$ and
$\omega_{\mathcal{GI}}=[\,\omega_{\mathcal I}^{\top}\ \omega_{\mathcal G}^{\top}\,]^{\top}$,
the frequency-divider relation~\eqref{eq:divider} follows. This formulation is used next to compose the system-level model leveraged for analysis. 

Collecting the device dynamics together with the algebraic power-balance relations yields the following semi-explicit DAE model:
\begin{subequations}
\label{eq:dae_single_area}
\begin{align}
\dot\theta_{\mathcal{GI}}
&=
I_{-1}\omega_{\mathcal{GI}}
\\
\dot\omega_{\mathcal{I}}
&=
-\,M_{\mathcal{I}}^{-1}D_{\mathcal{I}}\omega_{\mathcal{I}}
-\,M_{\mathcal{I}}^{-1}S_{\mathcal{I}}^\top\bar{\mathcal{B}}\theta \nonumber \\&\quad 
+\,M_{\mathcal{I}}^{-1}p^\mathrm{ref}_{\mathcal{I}}
-\,M_{\mathcal{I}}^{-1}p_{\ell,\mathcal{I}}
\\
\dot\omega_{\mathcal{G}}
&=
-\,M_{\mathcal{G}}^{-1}D_{\mathcal{G}}\omega_{\mathcal{G}}
-\,M_{\mathcal{G}}^{-1}S_{\mathcal{G}}^\top\bar{\mathcal{B}}\theta \nonumber \\
&\quad +\,M_{\mathcal{G}}^{-1}p_{m,\mathcal{G}}
-\,M_{\mathcal{G}}^{-1}p_{\ell,\mathcal{G}}
\\
\dot p_{m,\mathcal{G}}
&=
-\,T_{\mathcal{G}}^{-1}p_{m,\mathcal{G}}
-\,T_{\mathcal{G}}^{-1}R_{\mathcal{G}}^{-1}\omega_{\mathcal{G}}
+\,T_{\mathcal{G}}^{-1}p^\mathrm{ref}_{\mathcal{G}}
\label{eq:sg_governor_single_area}
\\
\zero
&=
-\,S_{\mathcal{L}}^\top\bar{\mathcal{B}}\theta
-\,p_{\ell,\mathcal{L}}
\\
\zero
&=
p^\mathrm{ref}_{\mathcal{F}}
-
K_{\mathcal{F}}\omega_{\mathcal{F}}
-
S_{\mathcal{F}}^\top\bar{\mathcal{B}}\theta
-
p_{\ell,\mathcal{F}}
\\
\zero
&=
\omega_{\mathcal{F}}
-
H_{\mathcal{I}}\omega_{\mathcal{I}}
-
H_{\mathcal{G}}\omega_{\mathcal{G}}
\end{align}
\end{subequations}
where the zero vectors are of appropriate dimensions, $
M_\mathcal{G}=\mathrm{diag}(M_i)_{i\in\mathcal{G}}$,
$M_\mathcal{I}=\mathrm{diag}(M_i)_{i\in\mathcal{I}}$,
$D_\mathcal{G}=\mathrm{diag}(D_i)_{i\in\mathcal{G}}$,
and $D_\mathcal{I}=\mathrm{diag}(D_i)_{i\in\mathcal{I}}$. For synchronous generators, parameters corresponding to the turbine dynamics are collected in $T_\mathcal{G}=\mathrm{diag}(\tau_i)_{i\in\mathcal{G}}$ and $R_\mathcal{G}=\mathrm{diag}(R_i)_{i\in\mathcal{G}}$. The frequency-sensitivity gains of the GFL IBRs are collected in $K_\mathcal{F}=\diag(k_{f,i})_{i\in\mathcal{F}}$.

The dynamic state vector collects the swing-bus angles, frequency deviations, and SG mechanical powers: $x =
\begin{bmatrix}
\theta_\cI^\top\ \ \theta_\cG^\top
\,
\;
\omega_\mathcal{I}^\top
\;
\omega_\mathcal{G}^\top
\;
p_{m,\mathcal{G}}^\top
\end{bmatrix}^{\!\top}
\in
\mathbb{R}^{\,3|\mathcal{G}|+2|\mathcal{I}|-1}$. On the other hand, the algebraic variables are: $z
=
\begin{bmatrix}
\theta_\mathcal{F}^\top
\;
\theta_\mathcal{L}^\top
\;
\omega_\mathcal{F}^\top
\end{bmatrix}^{\!\top}
\in
\mathbb{R}^{\,2|\mathcal{F}|+|\mathcal{L}|}$. Finally, the input vector to our model collects the active-power reference signals and load disturbances: $u
=
\begin{bmatrix}
(p^\mathrm{ref}_\mathcal{I})^\top
\;
(p^\mathrm{ref}_\mathcal{G})^\top
\;
(p^\mathrm{ref}_\mathcal{F})^\top
\;
p_\ell^\top
\end{bmatrix}^{\!\top}$.
With these definitions, we rewrite~\eqref{eq:dae_single_area} in the semi-explicit block form
\begin{subequations}
\label{eq:dae_block_augmented}
\begin{align}
\dot x &= A_{11}x + A_{12}z + B_{1}u, \label{eq:dae_block_augmented_x}
\\
\zero &= A_{21}x + A_{22}z + B_{2}u, \label{eq:dae_block_augmented_y} 
\end{align}
\end{subequations}
where
\begingroup
\small
\renewcommand{\arraystretch}{1.1}
\setlength{\arraycolsep}{2pt}
\begin{align*}
A_{11} &=
\begin{bmatrix}
\zero
& I_{-1,\cI}
& I_{-1,\cG}
& \zero
\\[1mm]
- M_\cI^{-1} S_\cI^\top \bar\cB \bar S_\mathcal{GI}
& - M_\cI^{-1} D_\cI
& \zero
& \zero
\\[1mm]
- M_\cG^{-1} S_\cG^\top \bar\cB \bar S_\mathcal{GI}
& \zero
& - M_\cG^{-1} D_\cG
& M_\cG^{-1}
\\[1mm]
\zero
& \zero
& - T_\cG^{-1} R_\cG^{-1}
& - T_\cG^{-1}
\end{bmatrix},
\\[6pt]
A_{12} &=
\begin{bmatrix}
\zero & \zero & \zero
\\[1mm]
- M_\cI^{-1} S_\cI^\top \bar\cB \bar S_\cF
& - M_\cI^{-1} S_\cI^\top \bar\cB \bar S_\cL
& \zero
\\[1mm]
- M_\cG^{-1} S_\cG^\top \bar\cB \bar S_\cF
& - M_\cG^{-1} S_\cG^\top \bar\cB \bar S_\cL
& \zero
\\[1mm]
\zero & \zero & \zero
\end{bmatrix},
\\[6pt]
A_{21} &=
\begin{bmatrix}
- S_\cF^\top \bar\cB \bar S_\mathcal{GI}
& \zero & \zero & \zero
\\[1mm]
- S_\cL^\top \bar\cB \bar S_\mathcal{GI}
& \zero & \zero & \zero
\\[1mm]
\zero
& - H_\cI
& - H_\cG
& \zero
\end{bmatrix}, \\[6pt]
A_{22} & =
\begin{bmatrix}
- S_\cF^\top \bar\cB \bar S_\cF
& - S_\cF^\top \bar\cB \bar S_\cL
& - K_\cF
\\[1mm]
- S_\cL^\top \bar\cB \bar S_\cF
& - S_\cL^\top \bar\cB \bar S_\cL
& \zero
\\[1mm]
\zero & \zero & I_{|\cF|}
\end{bmatrix},
\\[6pt]
B_{1} & =
\begin{bmatrix}
\zero & \zero & \zero & \zero
\\[1mm]
M_\cI^{-1} & \zero & \zero & -\,M_\cI^{-1} S_\cI^\top
\\[1mm]
\zero & \zero & \zero & -\,M_\cG^{-1} S_\cG^\top
\\[1mm]
\zero & T_\cG^{-1} & \zero & \zero
\end{bmatrix},
~~
B_{2} =
\begin{bmatrix}
\zero & \zero & I_{|\cF|} & -\,S_\cF^\top
\\[1mm]
\zero & \zero & \zero & -\,S_\cL^\top
\\[1mm]
\zero & \zero & \zero & \zero
\end{bmatrix}.
\end{align*}
\endgroup
Above, $I_{-1}$ is split as $I_{-1}=\big[\,I_{-1,\cI}\ \ I_{-1,\cG}\,\big]$.

\subsection{Reduction to ODE model} \label{sec:ODE_model}
To facilitate analysis we will reduce the model~\eqref{eq:dae_block_augmented} to an ODE. As a first step, we establish that the matrix $A_{22}$ is nonsingular. Recall that the algebraic state vector is $z=[\,\theta_\mathcal{LF}^\top\ \ \omega_\cF^\top\,]^\top$, where $\theta_\mathcal{LF}=[\,\theta_\cF^\top\ \theta_\cL^\top\,]^\top$. With this ordering, the algebraic block $A_{22}$ in
\eqref{eq:dae_block_augmented} has the partition
\[
A_{22}
=
\begin{bmatrix}
A^{\theta\theta}_{22} & A^{\theta\omega}_{22}\\[3pt]
A^{\omega\theta}_{22} & A^{\omega\omega}_{22}
\end{bmatrix}, ~~ A^{\theta\theta}_{22}
=
\begin{bmatrix}
- S_\cF^\top \bar\cB\,\bar S_\cF &
- S_\cF^\top \bar\cB\,\bar S_\cL
\\[3pt]
- S_\cL^\top \bar\cB\,\bar S_\cF &
- S_\cL^\top \bar\cB\,\bar S_\cL
\end{bmatrix}
\]
and note that $A^{\theta\theta}_{22}
=
-\,S_\mathcal{LF}^\top \bar\cB\,\bar S_\mathcal{LF}$. Recall that matrix $S_\mathcal{LF}^\top \bar\cB\,\bar S_\mathcal{LF}$ is symmetric positive definite. Consequently, $A^{\theta\theta}_{22} = -\,S_\mathcal{LF}^\top \bar\cB\,\bar S_\mathcal{LF} \prec 0$, and therefore $A^{\theta\theta}_{22}$ is nonsingular. The remaining blocks of $A_{22}$ are $A^{\omega\omega}_{22}=I_{|\cF|}$, $A^{\theta\omega}_{22}= [- K_\cF^\top\,\,\zero^\top]^\top$, $A^{\omega\theta}_{22}=\zero$. Hence $A_{22}$ is block upper triangular, and its determinant satisfies  $\det(A_{22}) = \det\!\big(A^{\theta\theta}_{22}\big) \times
\det\!\big(A^{\omega\omega}_{22}\big) = \det\!\big(A^{\theta\theta}_{22}\big)
\neq 0$. Therefore $A_{22}$ is nonsingular.

The algebraic constraint~\eqref{eq:dae_block_augmented_y} can be solved explicitly for the algebraic variables:
\begin{align}
z(t)
=
-\,A_{22}^{-1}\!\big(A_{21}x(t)+B_2u(t)\big) . 
\label{eq:z_solution}
\end{align}
Then, substituting the algebraic solution~\eqref{eq:z_solution}
into the differential equation~\eqref{eq:dae_block_augmented_x} yields the reduced ODE model: 
\begin{equation}
\label{eq:x_compact}
\dot x(t) = A x(t) + B u(t), \,\,\text{where}
\end{equation} 
\begin{align*}
    A = A_{11}-A_{12}A_{22}^{-1}A_{21}, \quad B = B_{1}-A_{12}A_{22}^{-1}B_{2} \, .
\end{align*}
The above reduction is obtained by taking the Schur complement with respect to the algebraic network variables, resulting in a purely dynamic model suitable for reachability analysis. In this formulation, reachability computations are carried out for the dynamic state vector $x$, while the algebraic variables $z$—representing the non-swing bus angles and GFL frequencies—can be subsequently reconstructed through $z = -A_{22}^{-1}(A_{21}x + B_2u)$.

\section{Reachability Method} \label{sec:reachability}
In this section, we present a computationally efficient framework for estimating reachable sets of linear control systems of the form~\eqref{eq:x_compact}. 
Given a set of possible inputs $\mathcal{U}$ and an initial set of states $\mathcal{X}_0$, we define the reachable set of the system~\eqref{eq:x_compact} for time $t\in[t_0,t_1]$ by:
\begin{align*}
\mathcal{R}_t(\mathcal{X}_0,\mathcal{U}) =
\{x(t)&: x(t) \text{ is a trajectory of}~\eqref{eq:x_compact}\,\,\text{for some} \\
u&:[t_0,t]\to\mathcal{U}\,\,\text{and}\,\,x(t_0)\in\mathcal{X}_0\}.
\end{align*}
In practice, the initial condition is obtained from the measured operating point of the network at time $t_0$, so that $\mathcal{X}_0$ typically reduces to a singleton. 

Recalling that $u=
\begin{bmatrix}
(p^\mathrm{ref}_\mathcal{I})^\top\;
(p^\mathrm{ref}_\mathcal{G})^\top\;
(p^\mathrm{ref}_\mathcal{F})^\top\;
p_\ell^\top
\end{bmatrix}^{\!\top}$,
the input set $\mathcal{U}$ captures uncertainty in net active-power demand $p_\ell$ as well as generator and inverter power references. This facilitates modeling, e.g., loss of load, loss of generation, and large load fluctuations. In the following, we consider an input set:
\[
\mathcal{U}=[\underline{u},\overline{u}]
=
\{u\in\mathbb{R}^{m}\mid \underline{u}_i\le u_i\le \overline{u}_i,\; i=1,\ldots,m\},
\]
with $m = |\mathcal I| + |\mathcal G| + |\mathcal F| + N$. In general, the input $u(t)$ may vary with time but is assumed to remain within $[\underline u,\overline u]$. Recall that computing the exact reachable set of~\eqref{eq:x_compact} is generally computationally challenging. Therefore, we seek computationally tractable \emph{over-approximations} that provably contain all possible trajectories of the system. Specifically, we compute a family of sets $\overline{\mathcal{R}}_t(\mathcal{X}_0,\mathcal{U})$ such that
\begin{align*}
 \mathcal{R}_t(\mathcal{X}_0,\mathcal{U}) \subseteq \overline{\mathcal{R}}_t(\mathcal{X}_0,\mathcal{U}), \forall t\in [t_0,t_1]. 
\end{align*}

\subsection{Over-Approximation of Reachable Sets} \label{sec:reach_transient}
Our approach separates the oscillatory and non-oscillatory modes of the linear system~\eqref{eq:x_compact} via the Jordan decomposition of matrix $A$, which enables a block-wise analysis of the dynamics. For non-oscillatory modes, we employ interval analysis to compute tight over-approximations of the reachable sets, while for oscillatory modes we use contraction theory to derive norm-ball over-approximations. The reachable set of the full system is then obtained by combining the reachable sets of each mode. We use the following fundamental result.

\begin{lemma} \label{lem:useful} Consider~\eqref{eq:x_compact} with the assumption that every eigenvalue of $A$ with nonzero imaginary part has the same algebraic and geometric multiplicity. Then, there exists a non-singular matrix $V\in\mathbb{R}^{n\times n}$ such that
\begin{equation}
\label{eq:Jdecomp}
V^{-1}AV =  
\left[\begin{smallmatrix}
J_{\mathrm{real}} & 0 & \cdots & 0\\
0 & J_2 & \cdots & 0\\
\vdots & \vdots & \ddots & \vdots\\
0 & 0 & \cdots & J_k
\end{smallmatrix}\right],
\end{equation}
where $J_{\mathrm{real}}\in \mathbb{R}^{\ell\times \ell}$ is a block-diagonal matrix consisting of all Jordan blocks associated with the real eigenvalues of $A$, and for $j\in\{ 2,\ldots,k\}$, $J_j =
\left[\begin{smallmatrix}
\alpha_j & \beta_j\\
-\beta_j & \alpha_j
\end{smallmatrix}\right]$ with $\alpha_j \pm i\beta_j$ being eigenvalues of $A$ with nonzero imaginary part. \hfill $\Box$
\end{lemma}

We refer to~\cite[Chapter 7]{CDM:01} for definition of algebraic and geometric multiplicity of an eigenvalue. The assumption on the eigenvalues of $A$ is not restrictive, since matrices with repeated complex eigenvalues form a measure-zero subset of all square matrices. This assumption is also verified numerically via case studies in Section~\ref{sec:simulations}. The proof of Lemma~\ref{lem:useful} then follows from the classical real Jordan decomposition theorem~\cite[Ch.~7]{CDM:01}. The significance of Lemma~\ref{lem:useful} lies in enabling a blockwise analysis of the dynamics in~\eqref{eq:x_compact}. In particular, consider the following coordinate transformation:
\[
y=
\begin{bmatrix}
y_{\mathrm{real}}^\top &
y_2^\top &
\cdots &
y_k^\top
\end{bmatrix}^{\!\top}
=V^{-1}x,
\]
where $y_{\mathrm{real}}\in\mathbb{R}^{\ell}$ corresponds to the Jordan block $J_{\mathrm{real}}$ and $y_j\in\mathbb{R}^{2}$ corresponds to the Jordan blocks $J_{j}$, for $j\in\{2,\ldots,k\}$. In this coordinate chart, $V^{-1}B =
\big[(V^{-1}B)_{\mathrm{real}},\,(V^{-1}B)_2,\ldots,(V^{-1}B)_k\big]$,
with $(V^{-1}B)_{\mathrm{real}}\in\mathbb{R}^{\ell\times m}$ and $(V^{-1}B)_j\in\mathbb{R}^{2\times m}$. Using this transformation, system~\eqref{eq:x_compact} decomposes into the \emph{non-oscillatory mode} dynamics:
\begin{align}
\label{eq:real}
\dot{y}_{\mathrm{real}}
=
J_{\mathrm{real}}y_{\mathrm{real}}
+
(V^{-1}B)_{\mathrm{real}}u,
\end{align}
and the \emph{oscillatory mode} dynamics:
\begin{align}
\label{eq:complex}
\dot{y}_j
=
J_j y_j +
(V^{-1}B)_j u,
\qquad j=2,\ldots,k.
\end{align}
Finally, we assume that the transformed initial set
$V^{-1}\mathcal{X}_0=\{V^{-1}x:\,x\in\mathcal{X}_0\}$
can be enclosed in a box, i.e.,
\begin{align}
\label{eq:initial-inclusion}
V^{-1}\mathcal{X}_0 \subseteq [\underline{y}_0,\overline{y}_0],
\end{align}
for some vectors $\underline{y}_0,\overline{y}_0\in\mathbb{R}^n$ with $\underline{y}_0\le \overline{y}_0$. 

Next, we develop tools to efficiently compute a tight over-approximation of the reachable sets of the oscillatory modes and the non-oscillatory modes of the power system~\eqref{eq:x_compact}. 

\subsubsection{Reachability of non-oscillatory modes}
We employ interval analysis to provide hyper-rectangular bounds on the trajectories of the non-oscillatory mode dynamics~\eqref{eq:real}. Using~\cite[Corollary 1]{jafarpour2024efficient}, we can construct the embedding system for~\eqref{eq:real} on $\mathbb{R}^{2\ell}$ as follows:
\begin{align}
\label{eq:embedding-real}
& \hspace{-.2cm}\dot{\underline{y}}_{\mathrm{real}}
=
J_{\mathrm{real}}\underline{y}_{\mathrm{real}}
+
[(V^{-1}B)_{\mathrm{real}}]^+\underline{u}
+
[(V^{-1}B)_{\mathrm{real}}]^-\overline{u},
\nonumber\\
& \hspace{-.2cm} \dot{\overline{y}}_{\mathrm{real}}
=
J_{\mathrm{real}}\overline{y}_{\mathrm{real}}
+
[(V^{-1}B)_{\mathrm{real}}]^+\overline{u}
+
[(V^{-1}B)_{\mathrm{real}}]^-\underline{u}.
\end{align}
Using~\cite[Proposition 5]{jafarpour2024efficient}, we can over-approximate the reachable set of~\eqref{eq:real} using a single trajectory of the embedding system~\eqref{eq:embedding-real}. This leads to the following result, whose proof is provided in Section~\ref{sec:proofs}.

\begin{theorem}[Reachability of non-oscillatory dynamics]
\label{thm:mm}
Consider non-oscillatory mode dynamics~\eqref{eq:real} with initial set $[(\underline{y}_0)_{\mathrm{real}},(\overline{y}_0)_{\mathrm{real}}]$, where $(\underline{y}_0)_{\mathrm{real}}\le (\overline{y}_0)_{\mathrm{real}}\in\mathbb{R}^\ell$ and inputs $u\in[\underline{u},\overline{u}]$. Let $t\mapsto
\left[\begin{smallmatrix}
\underline{y}_{\mathrm{real}}(t)\\
\overline{y}_{\mathrm{real}}(t)
\end{smallmatrix}\right]$ denote the trajectory of the embedding system~\eqref{eq:embedding-real} starting from $\left[\begin{smallmatrix}
(\underline{y}_0)_{\mathrm{real}}\\
(\overline{y}_0)_{\mathrm{real}}
\end{smallmatrix}\right]$. Then, for every $t\ge t_0$, $y_{\mathrm{real}}(t) \in [\underline{y}_{\mathrm{real}}(t),\overline{y}_{\mathrm{real}}(t)]$. \hfill $\Box$
\end{theorem}

\subsubsection{Reachability of oscillatory modes}
We use contraction theory to derive norm-ball bounds on the trajectories of the oscillatory mode dynamics~\eqref{eq:complex}. For notational simplicity, define $\underline g_j = [(V^{-1}B)_j]^+\underline u + [(V^{-1}B)_j]^-\overline u $, and $ \overline g_j = [(V^{-1}B)_j]^+\overline u + [(V^{-1}B)_j]^-\underline u$.

\begin{theorem}[Reachability of oscillatory dynamics]
\label{thm:contraction}
Consider the oscillatory mode dynamics~\eqref{eq:complex} with the initial set $[(\underline{y}_0)_{j},(\overline{y}_0)_{j}]$, where $(\underline{y}_0)_{j}\le (\overline{y}_0)_{j}\in\mathbb{R}^2$ and inputs $u\in[\underline u,\overline u]$. Then, for every $t\ge t_0$,
\begin{equation}
\label{eq:ball_complex}
y_j(t)\in \mathcal B\left(e^{\alpha_j(t-t_0)}r_{1,j}
+
\left(\tfrac{1-e^{\alpha_j(t-t_0)}}{-\alpha_j}\right)\,r_{2,j},\zero_2 \right),
\end{equation}
where 
\begin{align*}
r_{1,j} =
\left\|
\max\{|(\underline y_0)_j|,\;|(\overline y_0)_j|\}
\right\|_2,\,\,r_{2,j} =
\left\|
\max\{|\underline g_j|,\;|\overline g_j|\}
\right\|_2
\end{align*}
with maximum and absolute values taken component-wise. \hfill $\Box$
\end{theorem}

The proof is provided in Section~\ref{sec:proofs}. We combine Theorems~\ref{thm:mm} and~\ref{thm:contraction} to obtain an over-approximation of the reachable set of the dynamical model~\eqref{eq:x_compact}. 

\begin{theorem}[Reachable set over-approximations]
\label{thm:combined}
Consider the system~\eqref{eq:x_compact} with initial set $\mathcal X_0$ and inputs $u(t)\in[\underline u,\overline u]$ for all $t\ge t_0$. Let $\underline y_0\le \overline y_0$ satisfy~\eqref{eq:initial-inclusion}, and assume that, $\left[\begin{smallmatrix}
\underline y_{\mathrm{real}}(t)\\
\overline y_{\mathrm{real}}(t)
\end{smallmatrix}\right]$ is the trajectory of the embedding system~\eqref{eq:embedding-real} starting from $\left[\begin{smallmatrix}
(\underline y_0)_{\mathrm{real}}\\
(\overline y_0)_{\mathrm{real}}
\end{smallmatrix}\right]$. Then, for every $t\ge t_0$,
\begin{equation}
\mathcal R_t(\mathcal X_0,\mathcal U)
\subseteq
\overline{\mathcal R}^{\mathrm{dyn}}_t(\mathcal X_0,\mathcal U)
=
V\,\mathcal S_t,
\label{eq:combined_reach}
\end{equation}
where $\mathcal S_t
=
[\underline y_{\mathrm{real}}(t),\overline y_{\mathrm{real}}(t)]
\times
\mathcal B(\rho_2(t),\zero_2)
\times \cdots \times
\mathcal B(\rho_k(t),\zero_2)$,
with $\rho_j(t)$, $j=2,\ldots,k$ the radii defined in Theorem~\ref{thm:contraction}.
\hfill $\Box$
\end{theorem}

Although the set $\mathcal S_t$ is represented as the Cartesian product of one interval and multiple Euclidean balls, its image in the original state coordinates, $V\mathcal S_t$, is stored in the implementation through an interval enclosure $\underline{x}^{\mathrm{dyn}}(t),\overline{x}^{\mathrm{dyn}}(t)$, which facilitates efficient propagation. Once the set $\overline{\mathcal{R}}^{\mathrm{dyn}}_t(\mathcal{X}_0,\mathcal{U})$ of dynamic states is computed, an over-approximation of the reachable set of the algebraic variables $z$ can be reconstructed directly from the algebraic relation $z = -A_{22}^{-1}(A_{21}x + B_2u)$ by propagating the set $\overline{\mathcal{R}}^{\mathrm{dyn}}_t(\mathcal{X}_0,\mathcal{U})$ together with the admissible input set $\mathcal{U}$ through this mapping.

\subsection{Tightening tubes for piecewise-constant inputs}
\label{sec:reachability_trick}

In this section, we consider inputs $u$ that undergo an abrupt change of unknown magnitude at time $t_0$ and remain constant over the interval $[t_0,t_1]$. Such scenarios commonly arise in power systems following sudden changes in generation or load demand. By exploiting this piecewise-constant input structure, we derive tighter reachable-set bounds over $[t_0,t_1]$, reducing conservatism and enabling a more accurate characterization of transient behavior and settling times following the disturbance.

In this case, the linear system~\eqref{eq:x_compact} can be rewritten as a homogeneous system through an equilibrium shift. Since \(A\) is assumed nonsingular, define $x^\star(u)=-A^{-1}Bu$ and introduce the shifted state $\xi=x-x^\star(u)$,
which leads to the dynamics 
\begin{equation}
\label{eq:xi_dyn}
\dot \xi = A \xi .
\end{equation}
The uncertainty induced by the inputs now appears only through the initial condition $\xi(t_0) = x(t_0) - x^\star(u)$. Define the corresponding shifted initial set
\begin{equation}
\Xi_0 =
\{x_0 - x^\star(u): x_0\in\mathcal X_0,\; u\in[\underline u,\overline u]\}.
\end{equation}
The reachable set $\mathcal R_t^{\mathrm{hom}}(\Xi_0)$ associated with the homogeneous system~\eqref{eq:xi_dyn} admits the following exact characterization:
\[
\mathcal R^{\mathrm{hom}}_t(\Xi_0)
=
\{e^{A(t-t_0)}\xi_0:\xi_0\in\Xi_0\}.
\]
The following result shows that this reachable set can be bounded using the same modal reachability construction developed in Section~\ref{sec:reach_transient}.

\begin{theorem}[Reachable tube in shifted coordinates]
\label{thm:ss_tube}
Consider the system~\eqref{eq:x_compact} with initial set $\mathcal{X}_0$ and constant input $u \in [\underline{u},\overline{u}]$. Let $V^{-1}\Xi_0 \subseteq [\underline{\eta}_0,\overline{\eta}_0]$, where $V$ is the transformation matrix given in Lemma~\ref{lem:useful}. Then, for every $t\ge t_0$, 
\[
\mathcal R^{\mathrm{hom}}_t(\Xi_0)
\subseteq
\overline{\mathcal R}^{\mathrm{hom}}_t(\Xi_0)
=
V\mathcal S_\eta(t),
\]
where \(\mathcal S_\eta(t)\) is obtained by applying
Theorems~\ref{thm:mm}--\ref{thm:contraction} to the homogeneous system
\(\dot\eta = J\eta\) with initial set
\([\underline\eta_0,\overline\eta_0]\). \hfill $\Box$
\end{theorem}

Finally, since \(x=\xi+x^\star(u)\), the reachable set of the original state satisfies
\begin{align}
\mathcal R_t^{\mathrm{ss}}(\mathcal X_0,\mathcal U)
\subseteq
\overline{\mathcal R}_t^{\mathrm{ss}}(\mathcal X_0,\mathcal U)
:=
V\mathcal S_\eta(t)
\oplus
\mathcal X^\star(\mathcal U),
\end{align}
where $\mathcal X^\star(\mathcal U)
=
\{-A^{-1}Bu : u\in[\underline u,\overline u]\}$. In the proposed algorithm, this steady-state-informed tube is intersected with the dynamic reachable tube derived in Section~\ref{sec:reach_transient}, yielding tighter bounds on the system trajectories. Specifically, if the reachable tube from Theorem~\ref{thm:combined} at time $t$ is given by
$\overline{\mathcal R}_t^{\mathrm{dyn}}(\mathcal X_0,\mathcal U) = [\underline{x}^{\mathrm{dyn}}(t),\overline{x}^{\mathrm{dyn}}(t)]$ and the steady-state-informed tube from Theorem~\ref{thm:ss_tube} by $\overline{\mathcal R}_t^{\mathrm{ss}}(\mathcal X_0,\mathcal U) = [\underline{x}^{\mathrm{ss}}(t),\overline{x}^{\mathrm{ss}}(t)]$, then the fused reachable tube is computed component-wise as
\begin{align*}
\underline{x}(t) = \max\{\underline{x}^{\mathrm{dyn}}(t),\underline{x}^{\mathrm{ss}}(t)\}, \,\, \overline{x}(t) = \min\{\overline{x}^{\mathrm{dyn}}(t),\overline{x}^{\mathrm{ss}}(t)\}.
\end{align*}
Equivalently, $[\underline{x}(t),\overline{x}(t)]
=
\overline{\mathcal{R}}_t^{\mathrm{dyn}}(\mathcal{X}_0,\mathcal{U})\cap \overline{\mathcal R}^{\mathrm{ss}}_t(\mathcal{X}_0,\mathcal{U})$ and
\[
\mathcal R_t(\mathcal X_0,\mathcal U) \subseteq \overline{\mathcal R}_t(\mathcal{X}_0,\mathcal{U}) = 
[\underline{x}(t),\overline{x}(t)].
\]
The overall method is summarized as Algorithm~\ref{alg:reachability} below.

\begin{algorithm}[h!]
\caption{Reachability method for piecewise-constant input}
\label{alg:reachability}\footnotesize{
\begin{algorithmic}[1]

\State \textbf{Inputs:} time step $h>0$, event time $t_{\mathrm{event}}$, pre-event input $u^{\mathrm{pre}}\in\mathbb{R}^m$, post-event set $\mathcal U=[\underline u,\overline u]\subset\mathbb R^m$.
\State \textbf{Initialization:} $\mathcal R_0=\mathcal X_0=[\underline x_0,\overline x_0]$.

\For{$k=0,1,\ldots$ such that $t_0+kh<t_1$}
    \State $t_k=t_0+kh$
    
    \State \textbf{Input set at time $t_k$}
    \If{$t_k<t_{\mathrm{event}}$}
        \State $\mathcal U_k=\{u^{\mathrm{pre}}\}$
    \Else
        \State $\mathcal U_k=\mathcal U$
    \EndIf

    \State \textbf{Dynamic-state propagation}
    \State Compute $[\underline x^{\mathrm{dyn}}_{k+1},\overline x^{\mathrm{dyn}}_{k+1}]$ via Theorem~\ref{thm:combined}
    \If{$t_k\ge t_{\mathrm{event}}$}
        \State Compute $[\underline x^{\mathrm{ss}}_{k+1},\overline x^{\mathrm{ss}}_{k+1}]$ via Theorem~\ref{thm:ss_tube} 
        \State $\underline x_{k+1}=\max\{\underline x^{\mathrm{dyn}}_{k+1},\underline x^{\mathrm{ss}}_{k+1}\}$
        \State $\overline x_{k+1}=\min\{\overline x^{\mathrm{dyn}}_{k+1},\overline x^{\mathrm{ss}}_{k+1}\}$
    \Else
        \State $\underline x_{k+1}=\underline x^{\mathrm{dyn}}_{k+1}$
        \State $\overline x_{k+1}=\overline x^{\mathrm{dyn}}_{k+1}$
    \EndIf

    \State $\mathcal R_{k+1}=[\underline x_{k+1},\overline x_{k+1}]$
\EndFor

\end{algorithmic}}
\end{algorithm}

\section{Numerical Experiments and EMT Validation}
\label{sec:simulations}

\begin{figure}[t]
    \centering
    \includegraphics[width=0.7\columnwidth]{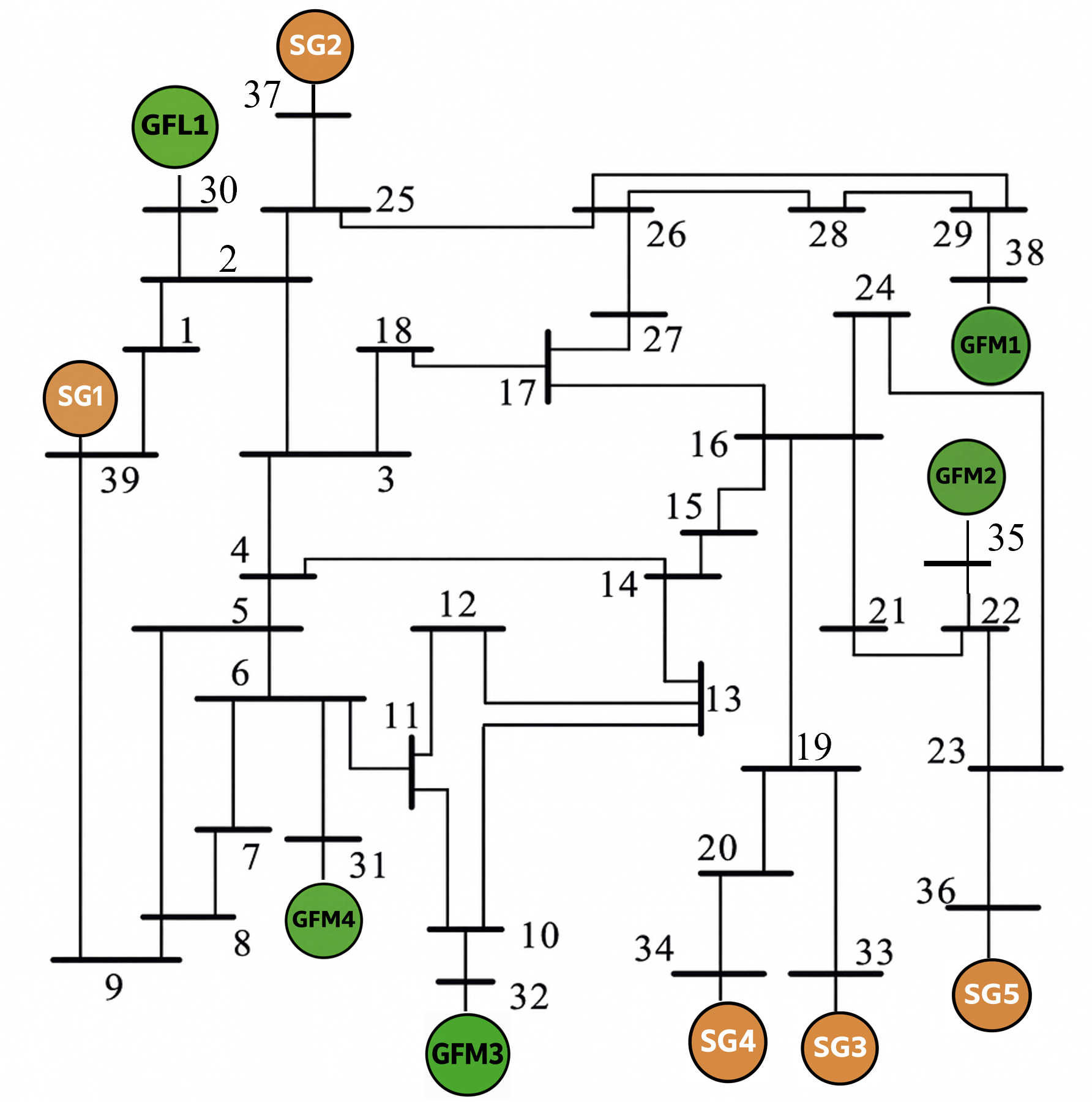}
    \caption{Modified IEEE New England 39-bus system with five SGs, four GFM IBRs, and one GFL IBR. Bus~39 is the angle reference bus.}
    \label{fig:system}
    \vspace{-.4cm}
\end{figure}

We consider a modified IEEE 39-bus New England test system on a 100~MVA base at 60~Hz, comprising 39 buses, 10 generator buses, and 46 transmission branches and transformers. As shown in Fig.~\ref{fig:system}, the system includes five synchronous generators (SGs), four grid-forming (GFM) inverters, and one grid-following (GFL) inverter.

To validate the analytical model used in the reachability analysis, we perform high-fidelity EMT simulations of the IEEE 39-bus system in \texttt{PowerSimulationsDynamics.jl}~(PSID)~\cite{lara2023powersimulationsdynamics}. PSID is an open-source Julia framework for quasi-static phasor~(QSP) and $dq0$ EMT time-domain simulation. It assembles device-level equations into a unified DAE model and can capture electromagnetic, electromechanical, network, and converter-control dynamics~\cite{lara2023powersimulationsdynamics}. The EMT models considered here are summarized in Table~\ref{tab:emt_vs_analytical}. Since the resulting DAE is stiff due to the separation between electromechanical ($\mathcal{O}(1~\mathrm{s})$) and converter-control ($\mathcal{O}(1~\mathrm{ms})$) time scales~\cite{Lara-2024}, we formulate the simulations using PSID's \texttt{ResidualModel}, which expresses the equations in implicit residual form and solves them with the \textsc{SUNDIALS} \texttt{IDA()} solver~\cite{hindmarsh2005sundials}.

\subsubsection{System Initialization}
\label{sec:psid_init}

The modified IEEE 39-bus system is specified in MATPOWER format and imported into a \texttt{PowerSystems.jl} \texttt{System} object~\cite{lara2021powersystems}. Network quantities are expressed on the system base, while device parameters follow the equipment MVA bases under the \texttt{PowerSystems.jl} per-unit convention. The base operating point is obtained from a full AC Newton–Raphson power flow solved using \texttt{PowerFlows.jl}. The converged bus voltages and phase angles are written back to the \texttt{System} object and used to initialize all dynamic models. 

\subsubsection{Device Models for EMT Validation}
\label{sec:psid_components}

Device models are summarized in Tables~\ref{tab:emt_vs_analytical} and~\ref{tab:ieee39_models_psid}. Each synchronous generator is modeled using a round-rotor machine representation~\cite{sauer2017power} with transient internal voltages on the $d$ and $q$ axes and subtransient rotor flux states. The model includes a single-mass shaft representation for rotor dynamics, a Type~1 AVR excitation system for terminal-voltage regulation, a turbine-governor for primary frequency control, and a power system stabilizer for damping electromechanical oscillations.

The inverter-based resources comprise GFM and GFL converters. Within PSID, each inverter is represented by a DC source, a switching-stage model, an output-side $LCL$ filter, a fast inner control loop, an outer control loop, and a PLL-based frequency estimator. The switching stage is modeled through an averaged inverter representation, where modulation signals determine the AC-side $dq$-frame voltage generated from the DC source~\cite{yazdani2010voltage}. The $LCL$ filter captures the interface with the grid. For GFM units, the outer control loop emulates a virtual synchronous machine with virtual inertia and reactive-power droop, while the inner loop regulates converter voltage in the synchronous reference frame. For GFL units, synchronization is achieved through a PLL, and the outer control loop regulates active and reactive power through PI controllers that generate $dq$ current references, which are tracked by the fast inner current loop in the PLL-defined synchronous reference frame.

\begin{table}[t]
\centering
\caption{Models used in IEEE-39 dq-EMT simulations in \texttt{PowerSimulationsDynamics.jl}.}
\label{tab:ieee39_models_psid}

\renewcommand{\arraystretch}{1.06}
\setlength{\tabcolsep}{3pt}
\scriptsize

\resizebox{\columnwidth}{!}{%
\begin{tabular}{c c c c c c c}
\hline
\textbf{Gens} & \textbf{Machine} & \textbf{Shaft} & \textbf{Excitation} & \textbf{Governor} & \textbf{PSS} & \\
\hline
\textbf{SG} &
\texttt{RoundRotorQuadratic} &
\texttt{SingleMass} &
\texttt{AVRTypeI} &
\texttt{TGTypeI} &
\texttt{IEEEST} &
\\
\hline
\textbf{IBR} & \textbf{Outer} & \textbf{Inner} & \textbf{Converter} & \textbf{Filter} & \textbf{Freq. Est.} & \\
\hline
\textbf{GFM} &
\begin{tabular}[c]{@{}c@{}}
\texttt{VirtualInertia}\\
\texttt{ReactivePowerDroop}
\end{tabular} &
\texttt{VoltageModeControl} &
\texttt{AverageConverter} &
\texttt{LCLFilter} &
None &
\\
\hline
\textbf{GFL} &
\begin{tabular}[c]{@{}c@{}}
\texttt{ActivePowerPI}\\
\texttt{ReactivePowerPI}
\end{tabular} &
\texttt{CurrentModeControl} &
\texttt{AverageConverter} &
\texttt{LCLFilter} &
\texttt{KauraPLL} &
\\
\hline
\end{tabular}%
}
\vspace{-.4cm}
\end{table}

\subsubsection{Network and Load Representation}
\label{sec:psid_network}

Transmission lines are modeled in PSID using \texttt{DynamicBranch} dynamic-$\pi$ models, which introduce branch dynamic states into the dq0 differential-algebraic formulation and capture line transient dynamics. Loads are represented using ZIP models, with only the constant-power component considered. Disturbances are introduced through PSID perturbation objects.

\vspace{-0.2cm}
\subsection{Scenarios}
We consider two representative scenarios; however, the proposed method applies to a broader class of disturbance scenarios that are not included here due to space constraints.

\begin{figure*}[t]
    \centering
    \includegraphics[width=1.0\textwidth,keepaspectratio]{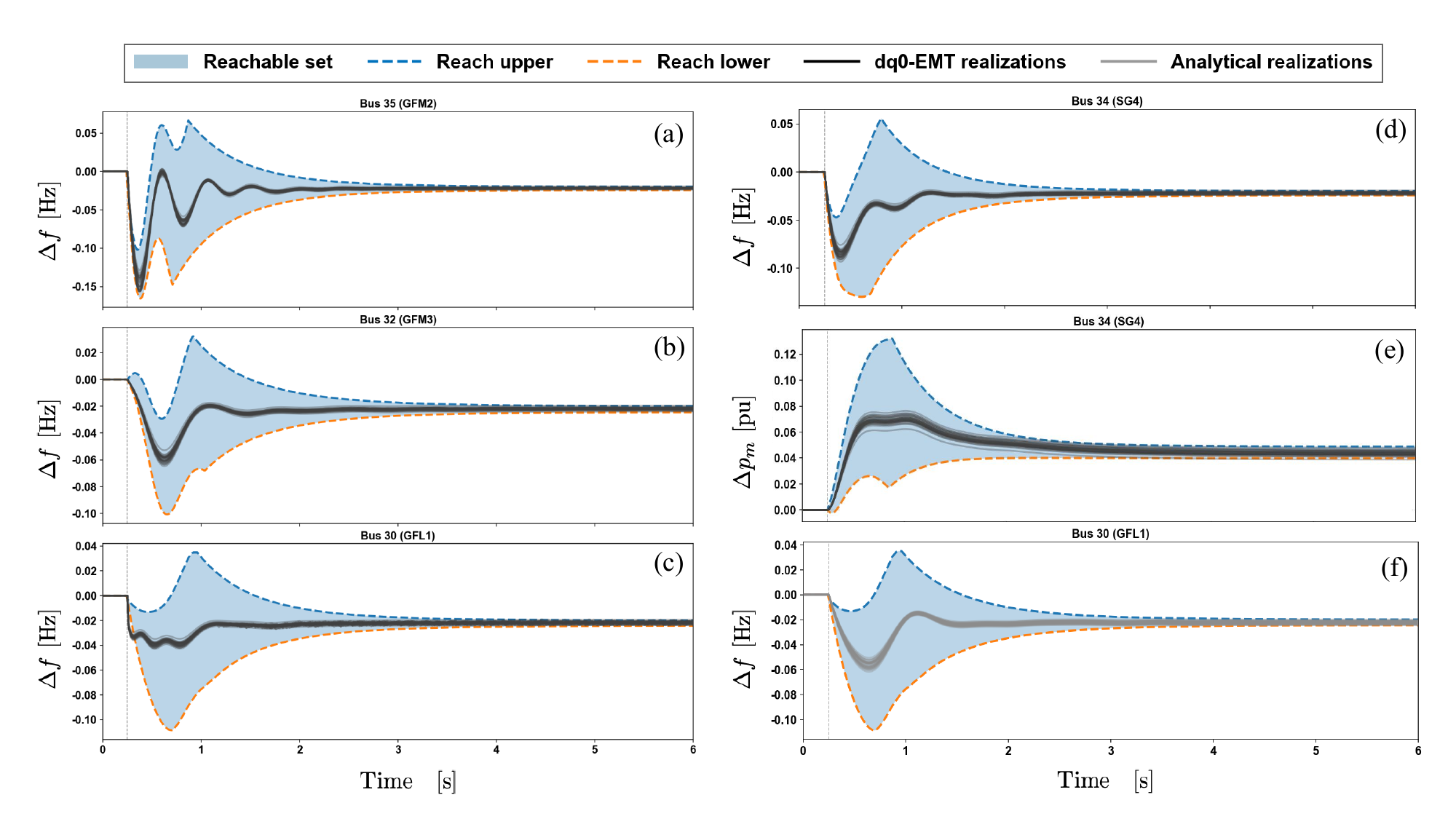}
    \caption{Reachable sets and trajectories from the EMT simulations for Scenario 1, for representative buses. The vertical line indicates the time of the load event. Plots show frequency deviations for two GFM IBR buses in (a)-(b), one GFL IBR bus in (c), and one generator bus in (d). Panel (e) shows the power deviation of the generator at bus 34. Panels (c) and (f) compare the trajectories obtained using the EMT model and the analytical model~\eqref{eq:x_compact}. }
    \label{fig:scenario1}
    \vspace{-.4cm}
\end{figure*}

$\triangle$ \emph{\textbf{Scenario 1: Concurrent load events at two buses}}. We first consider load steps applied concurrently at two buses. Disturbances are introduced at bus~35 and bus~20 at time $t_{\mathrm{event}}=0.25~\mathrm{s}$, with uncertain load variations modeled as intervals $\mathcal{U}_{35}=[1.8,2.2]$~p.u.\ and $\mathcal{U}_{20}=[1.35,1.65]$~p.u., respectively. The reachable sets are computed over the time horizon $[t_0,t_1]=[0,6]~\mathrm{s}$ using the analytical model developed in this paper. To assess the accuracy of the reachability approximation, we generate 30 load realizations within these uncertainty bounds and simulate the corresponding system trajectories using the high-fidelity $dq0$ EMT model implemented in PSID.\ Figure~\ref{fig:scenario1} compares the resulting reachable sets with the EMT trajectories, for representative IBRs and SGs. The EMT trajectories remain largely contained within the computed reachable tubes, demonstrating that the proposed method provides a tight over-approximation of the system dynamics under bounded disturbances. A small discrepancy is observed at the GFL bus immediately after the disturbance, where the EMT frequency trajectory briefly exits the reachable tube for approximately $0.1~\mathrm{s}$. This deviation is expected and arises from the frequency-divider approximation used to express the GFL frequency as a function of the swing-bus frequencies. Since this approximation neglects the fast PLL dynamics, the mismatch is confined to a short time interval and quickly disappears as the system evolves. 

\begin{figure*}[t]
    \centering
\includegraphics[width=1.0\textwidth,keepaspectratio]{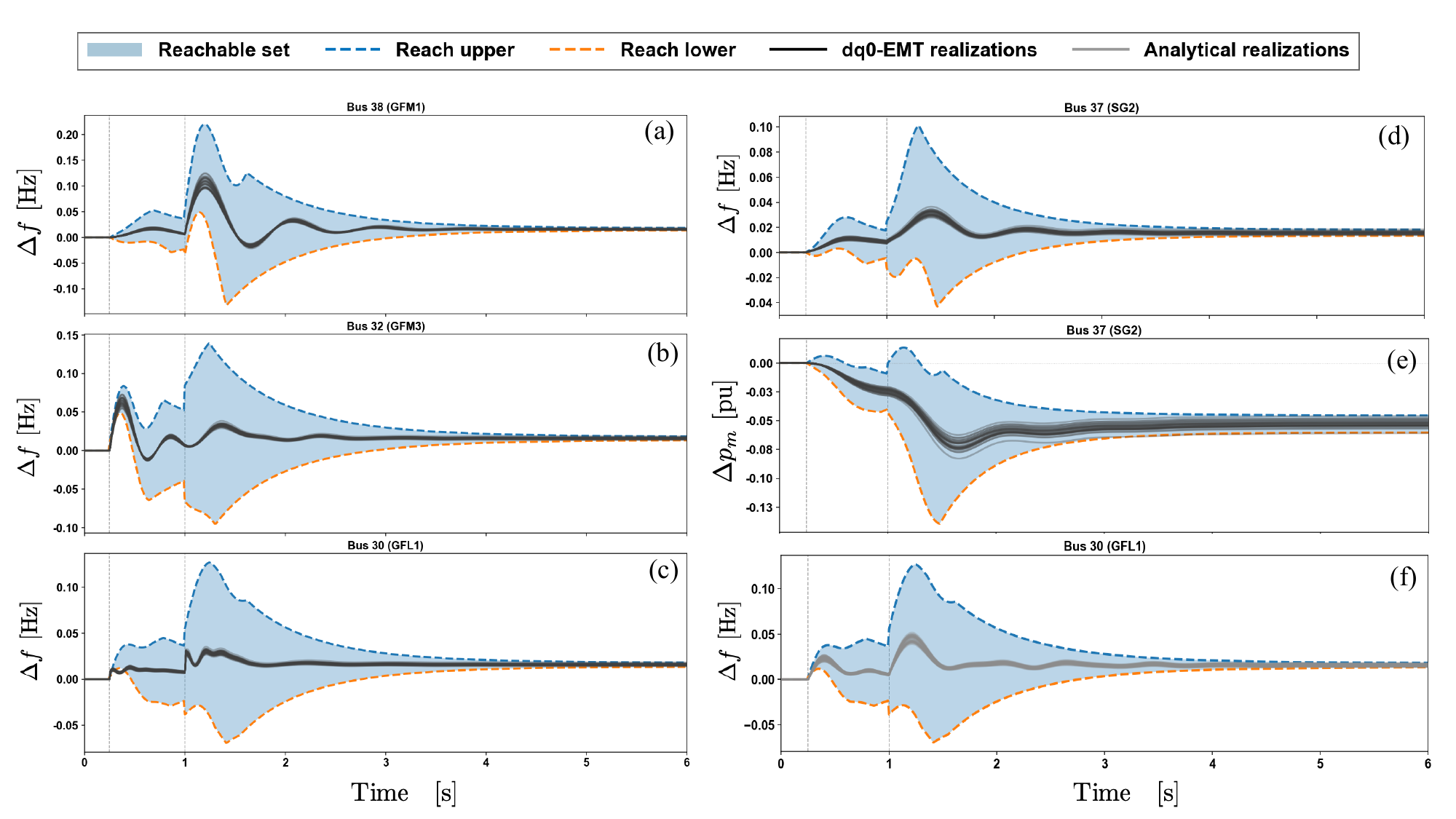}
    \caption{Reachable-set over-approximations and EMT-simulation trajectories for Scenario 2, for representative buses. Vertical lines indicate the load events.}
    \label{fig:scenario2}
    \vspace{-.4cm}
\end{figure*}

$\triangle$ \emph{\textbf{Scenario 2: Consecutive events with net load variation}}. This scenario illustrates the ability of the proposed reachability framework to track system evolution under multiple sequential disturbances, a situation that would require numerous time-domain simulations using conventional analysis methods. We consider consecutive disturbance events producing a net load variation over time. The first disturbance is applied at bus~32 at time $t_{\mathrm{event}}=0.25~\mathrm{s}$, with uncertain load variation $\mathcal{U}_{32}=[-1.15,-0.85]$~p.u. A second disturbance occurs at bus~38 at time $t_{\mathrm{event}}=1.0~\mathrm{s}$ with uncertainty $\mathcal{U}_{38}=[-1.70,-1.30]$~p.u.

Because the disturbances occur sequentially, the reachability analysis generates two consecutive reachable tubes corresponding to the system evolution before and after the second event. To validate the results, thirty $dq0$ EMT trajectories are generated using PSID by sampling disturbance realizations within the specified uncertainty sets. As shown in Fig.~\ref{fig:scenario2}, the close agreement between the EMT trajectories and the computed reachable tubes confirms that the proposed framework captures the system dynamics with high accuracy.

\begin{table}[t]
\centering
\caption{Runtime comparison for \(N_s=30\) realizations drawn from \(\mathcal U\).}
\label{tab:runtime_comparison}
\renewcommand{\arraystretch}{1.0}
\setlength{\tabcolsep}{7pt}
\begin{tabular}{lrr}
\hline
\textbf{Method} & \textbf{Runtime [s]} & \textbf{Relative} \\
\hline
Proposed method & 0.0956 & $1\times$ \\
Trajectory roll-out using~\eqref{eq:x_compact} & 0.3812 & $4.0\times$ \\
dq0-EMT  & 384.1110 & $4.0\times10^{3}\times$ \\
\hline
\end{tabular}
\vspace{-.4cm}
\end{table}

\vspace{-0.2cm}
\subsection{Computational Cost of Reachability Algorithm}
We compare the computational times of the proposed method with the time required to generate 30 trajectories starting from 30 different values in $\mathcal{U}$.  Table~\ref{tab:runtime_comparison} compares the computational times: the reachable tube is computed in $0.0956$~s, whereas generating $N_s=30$ trajectories requires $0.3812$~s using the reduced model and $384.1110$~s for the dq0-EMT simulations. The proposed reachability method is approximately $4\times$ faster than computing 30 analytical trajectories and about $4.0\times10^{3}$ times faster than the EMT simulations used for validation. The reported computation times were obtained using a workstation equipped with a 16-core Intel Core Ultra 7 255H processor and 32~GB of RAM.

\section{Conclusions}
We presented a computationally efficient framework for reachability analysis of power-system dynamics. The method yields guaranteed over-approximations of reachable sets under bounded load and generation variations. Numerical results on the IEEE 39-bus system showed that reachable tubes can be computed with very low computational cost: multi-second reachable sets were computed in about $0.1~\mathrm{s}$, making the approach faster than repeated trajectory roll-outs and several orders of magnitude faster than high-fidelity $dq0$ EMT simulations.

\section{Proofs} \label{sec:proofs}
\subsection{Invertibility of $S_{\mathcal{LF}}^\top\bar{\mathcal{B}}\bar S_{\mathcal{LF}}$} \label{sec:InvertibilityForFD}
Partition the buses as $(\bar{\mathcal N},\{s\})$, where
$\bar{\mathcal N}=\{1,\ldots,N-1\}$ denotes the non-angle-reference buses.
With this ordering, the Laplacian admits the decomposition $\mathcal B=
\begin{bmatrix}
\mathcal B_{\mathrm{ground}} & \mathbf b_{gs}\\
\mathbf b_{sg}^\top & b_{ss}
\end{bmatrix}$, 
where $\mathcal B_{\mathrm{ground}}\in\mathbb R^{(N-1)\times(N-1)}$
is the grounded Laplacian obtained by removing the angle-reference bus row and column.
Moreover, $\bar{\mathcal B}
=
\mathcal B I_0
=
\begin{bmatrix}
\mathcal B_{\mathrm{ground}}\\
\mathbf b_{sg}^\top
\end{bmatrix}$. Since the angle-reference bus is not in $\mathcal{LF}$, the selector satisfies $S_{\mathcal{LF}}=
\begin{bmatrix}
\bar S_{\mathcal{LF}}\\
\zero_{1\times|\mathcal{LF}|}
\end{bmatrix}$. Therefore, $S_{\mathcal{LF}}^\top\bar{\mathcal B}\bar S_{\mathcal{LF}}
=
\bar S_{\mathcal{LF}}^\top
\mathcal B_{\mathrm{ground}}
\bar S_{\mathcal{LF}}$. The grounded Laplacian $\mathcal B_{\mathrm{ground}}$ is symmetric positive definite since the network graph is connected~\cite[Prop.~5.1]{dorfler2018electrical}. Since every principal submatrix
of a positive definite matrix is positive definite
\cite[Sec.~4.2.1]{golub2013matrix}, the matrix
\(
\bar S_{\mathcal{LF}}^\top
\mathcal B_{\mathrm{ground}}
\bar S_{\mathcal{LF}}
\)
is symmetric positive definite. Hence
\(S_{\mathcal{LF}}^\top\bar{\mathcal B}\bar S_{\mathcal{LF}}\)
is invertible, which ensures that the expression defining $H$
is well-defined.

\subsection{Proof of Theorem~\ref{thm:mm}}

Define the function
\begin{small}
\[
\mathcal{F}(\underline{y}_{\mathrm{real}},\overline{y}_{\mathrm{real}},\underline{u},\overline{u})
=
\begin{bmatrix}
J_{\mathrm{real}}\underline{y}_{\mathrm{real}}
+
[(G)_{\mathrm{real}}]^+\underline{u}
+
[(G)_{\mathrm{real}}]^-\overline{u}
\\
J_{\mathrm{real}}\overline{y}_{\mathrm{real}}
+
[(G)_{\mathrm{real}}]^+\overline{u}
+
[(G)_{\mathrm{real}}]^-\underline{u}
\end{bmatrix}
\]
\end{small}
where $G = V^{-1}B$. For every $\underline{y}_{\mathrm{real}}\le y\le\overline{y}_{\mathrm{real}}$ and every $\underline{u}\le u\le\overline{u}$ we have
\begin{multline*}
J_{\mathrm{real}}\underline{y}_{\mathrm{real}}
+
[(G)_{\mathrm{real}}]^+\underline{u}
+
[(G)_{\mathrm{real}}]^-\overline{u}
\le
J_{\mathrm{real}}y
+
(G)_{\mathrm{real}}u
\\
\le
J_{\mathrm{real}}\overline{y}_{\mathrm{real}}
+
[(G)_{\mathrm{real}}]^+\overline{u}
+
[(G)_{\mathrm{real}}]^-\underline{u}.
\end{multline*}
This implies that $\mathcal{F}$ is an inclusion function for the real Jordan dynamics~\eqref{eq:real} and the result follows from~\cite[Proposition 5]{jafarpour2024efficient}.

\subsection{Proof of Theorem~\ref{thm:contraction}}

Consider the subsystem $\dot y_j = J_j y_j + G_j u$, $J_j=
\left[\begin{smallmatrix}
\alpha_j & \beta_j\\
-\beta_j & \alpha_j
\end{smallmatrix}\right]$. We first note that $J_j=\alpha_j I_2+\beta_j
\left[\begin{smallmatrix}
0 & 1\\
-1 & 0
\end{smallmatrix}\right]$ and therefore, $J_j+J_j^\top = 2\alpha_j I_2$. Hence, the $\ell_2$ logarithmic norm of $J_j$ is given by $
\mu_2(J_j)
= \lambda_{\max}\!(\frac{J_j+J_j^\top}{2}) =
\alpha_j$~\cite[Table 2.1]{FB:24-CTDS}.

Let $u(\cdot)$ be any measurable input satisfying
$u(\tau)\in[\underline u,\overline u]$ for all $\tau\ge t_0$, and let
$y_j(\cdot)$ be the corresponding solution. By~\cite[Theorem 37]{davydov2022non}, $\frac{d}{dt}\|y_j(t)\|_2
\le
\mu_2(J_j)\|y_j(t)\|_2+\|G_j u(t)\|_2
=
\alpha_j\|y_j(t)\|_2+\|G_j u(t)\|_2$.
Applying Gr\"onwall's inequality on the interval $[t_0,t]$ gives
\[
\|y_j(t)\|_2
\le
e^{\alpha_j(t-t_0)}\|y_j(t_0)\|_2
+
\int_{t_0}^t e^{\alpha_j(t-\tau)}\|G_j u(\tau)\|_2\,d\tau .
\]

Next, since $y_j(t_0)\in[(\underline y_0)_j,(\overline y_0)_j]$, it follows that $|y_j(t_0)|
\le
\max\Big\{|(\underline y_0)_j|,\;|(\overline y_0)_j|\Big\}$. Therefore,
$\|y_j(t_0)\|_2
\le
\left\|
\max\Big\{|(\underline y_0)_j|,\;|(\overline y_0)_j|\Big\}
\right\|_2
=
r_{1,j}$.

Next, for every $u\in[\underline u,\overline u]$, we have that $
G_j u \in [\underline g_j,\overline g_j]$,
where $\underline g_j=[G_j]^+\underline u+[G_j]^-\overline u$ and $\overline g_j=[G_j]^+\overline u+[G_j]^-\underline u$.
Hence, componentwise, $|G_j u|
\le
\max\{|\underline g_j|,\;|\overline g_j|\}$,
and thus $\|G_j u\|_2
\le
\left\|
\max\{|\underline g_j|,\;|\overline g_j|\}
\right\|_2
=
r_{2,j}$. Since this bound holds for every input $u(\tau)$, we obtain
\[
\|y_j(t)\|_2
\le
e^{\alpha_j(t-t_0)}r_{1,j}
+
r_{2,j}\int_{t_0}^t e^{\alpha_j(t-\tau)}\,d\tau .
\]
Because $\alpha_j<0$, we have that $
\|y_j(t)\|_2
\le
e^{\alpha_j(t-t_0)}r_{1,j}
+
\frac{1-e^{\alpha_j(t-t_0)}}{-\alpha_j}\,r_{2,j}$. Since this bound holds for every admissible trajectory, it follows that $y_j(t)\in
\mathcal{B}\bigl(\rho_j(t),\zero_2\bigr)$, $\rho_j(t)
=
e^{\alpha_j(t-t_0)}r_{1,j}
+
\frac{1-e^{\alpha_j(t-t_0)}}{-\alpha_j}\,r_{2,j}$, which proves the claim.

\bibliographystyle{IEEEtran}
\bibliography{biblio}

@INPROCEEDINGS{Choi-CDC-2015,
  author={Choi, Hyungjin and Seiler, Peter J. and Dhople, Sairaj V.},
  booktitle={2015 54th IEEE Conference on Decision and Control (CDC)}, 
  title={Uncertainty propagation with Semidefinite Programming}, 
  year={2015},
  volume={},
  number={},
  pages={5966-5971},
  doi={10.1109/CDC.2015.7403157}}

@INPROCEEDINGS{Dhople-SetTheoretic-2016,
  author={Al-Digs, Abdullah and Dhople, Sairaj V. and Chen, Yu Christine},
  booktitle={2016 IEEE Power \& Energy Society Innovative Smart Grid Technologies Conference (ISGT)}, 
  title={Estimating feasible nodal power injections in distribution networks}, 
  year={2016},
  volume={},
  number={},
  pages={1-5},
  doi={10.1109/ISGT.2016.7781244}}

@ARTICLE{ADG-2012,
  author={Chen, Yu Christine and Dominguez-Garcia, Alejandro D.},
  journal={IEEE Transactions on Power Systems}, 
  title={A Method to Study the Effect of Renewable Resource Variability on Power System Dynamics}, 
  year={2012},
  volume={27},
  number={4},
  pages={1978-1989},
  doi={10.1109/TPWRS.2012.2194168}}

@ARTICLE{Suul-Equivalence-2014,
  author={D'Arco, Salvatore and Suul, Jon Are},
  journal={IEEE Transactions on Smart Grid}, 
  title={Equivalence of Virtual Synchronous Machines and Frequency-Droops for Converter-Based MicroGrids}, 
  year={2014},
  volume={5},
  number={1},
  pages={394-395},
  doi={10.1109/TSG.2013.2288000}}

@article{TAYLOR20161322,
title = {Power systems without fuel},
journal = {Renewable and Sustainable Energy Reviews},
volume = {57},
pages = {1322-1336},
year = {2016},
issn = {1364-0321},
author = {Josh A. Taylor and Sairaj V. Dhople and Duncan S. Callaway}
}

@misc{Venkat-2022,
      title={Integrated System Models for Networks with Generators \& Inverters}, 
      author={D. Venkatramanan and Manish K. Singh and Olaolu Ajala and Alejandro Dominguez-Garcia and Sairaj Dhople},
      year={2022},
      eprint={2203.08253},
      archivePrefix={arXiv},
      primaryClass={eess.SY},
      url={https://arxiv.org/abs/2203.08253}, 
}

@ARTICLE{Lara-2024,
  author={Lara, Jose Daniel and Henriquez-Auba, Rodrigo and Ramasubramanian, Deepak and Dhople, Sairaj and Callaway, Duncan S. and Sanders, Seth},
  journal={IEEE Transactions on Power Systems}, 
  title={Revisiting Power Systems Time-Domain Simulation Methods and Models}, 
  year={2024},
  volume={39},
  number={2},
  pages={2421-2437},
  doi={10.1109/TPWRS.2023.3303291}}

@ARTICLE{Choi-2017,
  author={Choi, Hyungjin and Seiler, Peter J. and Dhople, Sairaj V.},
  journal={IEEE Transactions on Power Systems}, 
  title={{Propagating Uncertainty in Power-System DAE Models With Semidefinite Programming}}, 
  year={2017},
  volume={32},
  number={4},
  pages={3146-3156},
  doi={10.1109/TPWRS.2016.2615600}}

@article{jafarpour2024efficient,
  title={Efficient interaction-aware interval analysis of neural network feedback loops},
  author={Jafarpour, Saber and Harapanahalli, Akash and Coogan, Samuel},
  journal={IEEE Transactions on Automatic Control},
  volume={69},
  number={12},
  pages={8706--8721},
  year={2024},
  publisher={IEEE}
}

@article{davydov2022non,
  title={Non-Euclidean contraction theory for robust nonlinear stability},
  author={Davydov, Alexander and Jafarpour, Saber and Bullo, Francesco},
  journal={IEEE Transactions on Automatic Control},
  volume={67},
  number={12},
  pages={6667--6681},
  year={2022},
  publisher={IEEE}
}

@Book{FB:24-CTDS,
  author =	 {F. Bullo},
  title =	 {Contraction Theory for Dynamical Systems},
  year =	 2024,
  edition =	 {{1.2}},
  publisher =	 {Kindle Direct Publishing},
  ISBN =	 {979-8836646806}
}

@inproceedings{bansal2017hamilton,
  title={Hamilton-{Jacobi} reachability: A brief overview and recent advances},
  author={Bansal, Somil and Chen, Mo and Herbert, Sylvia and Tomlin, Claire J},
  booktitle={IEEE Conference on Decision and Control},
  pages={2242--2253},
  year={2017},
  organization={IEEE}
}

@article{ajala2020library,
  title={A library of second-order models for synchronous machines},
  author={Ajala, Olaoluwapo and Dom{\'\i}nguez-Garc{\'\i}a, Alejandro and Sauer, Peter and Liberzon, Daniel},
  journal={IEEE Transactions on Power Systems},
  volume={35},
  number={6},
  pages={4803--4814},
  year={2020},
  publisher={IEEE}
}

@article{milano2016frequency,
  title={Frequency divider},
  author={Milano, Federico and Ortega, Alvaro},
  journal={IEEE Transactions on Power Systems},
  volume={32},
  number={2},
  pages={1493--1501},
  year={2016},
  publisher={IEEE}
}

@inproceedings{milano2018foundations,
  title={Foundations and challenges of low-inertia systems},
  author={Milano, Federico and D{\"o}rfler, Florian and Hug, Gabriela and Hill, David J and Verbi{\v{c}}, Gregor},
  booktitle={Power Systems Computation Conference},
  year={2018}
}

@article{dorfler2023control,
  title={Control of low-inertia power systems},
  author={D{\"o}rfler, Florian and Gro{\ss}, Dominic},
  journal={Annual Review of Control, Robotics, and Autonomous Systems},
  volume={6},
  number={1},
  pages={415--445},
  year={2023},
  publisher={Annual Reviews}
}

@article{dorfler2018electrical,
  title={Electrical networks and algebraic graph theory: Models, properties, and applications},
  author={D{\"o}rfler, Florian and Simpson-Porco, John W and Bullo, Francesco},
  journal={Proceedings of the IEEE},
  volume={106},
  number={5},
  pages={977--1005},
  year={2018},
  publisher={IEEE}
}

@book{golub2013matrix,
  title={Matrix computations},
  author={Golub, Gene H and Van Loan, Charles F},
  year={2013},
  publisher={JHU press}
}

@inproceedings{pico2013reachability,
  title={Reachability analysis of power system frequency dynamics with new high-capacity HVAC and HVDC transmission lines},
  author={Pico, Hugo N Villegas and Aliprantis, Dionysios C and Hoff, Elena C},
  booktitle={2013 IREP Symposium Bulk Power System Dynamics and Control-IX Optimization, Security and Control of the Emerging Power Grid},
  pages={1--9},
  year={2013},
  organization={IEEE}
}

@article{althoff2014formal,
  title={Formal and compositional analysis of power systems using reachable sets},
  author={Althoff, Matthias},
  journal={IEEE Transactions on Power Systems},
  volume={29},
  number={5},
  pages={2270--2280},
  year={2014},
  publisher={IEEE}
}

@Book{CDM:01,
  author =	 {C. D. Meyer},
  title =	 {Matrix Analysis and Applied Linear Algebra},
  publisher =	{Society for Industrial and Applied Mathematics},
  year =	 {2013},
  address = {Philadelphia, PA},
  doi = {10.1137/1.9781611977448},
}

@article{jin2010reachability,
  title={Reachability analysis based transient stability design in power systems},
  author={Jin, Licheng and Kumar, Ratnesh and Elia, Nicola},
  journal={International Journal of Electrical Power \& Energy Systems},
  volume={32},
  number={7},
  pages={782--787},
  year={2010},
  publisher={Elsevier}
}

@inproceedings{el2017compositional,
  title={Compositional transient stability analysis of power systems via the computation of reachable sets},
  author={El-Guindy, Ahmed and Chen, Yu Christine and Althoff, Matthias},
  booktitle={American Control Conference},
  pages={2536--2543},
  year={2017},
  organization={IEEE}
}

@article{kundur2007power,
  title={Power system stability},
  author={Kundur, Prabha and others},
  journal={Power system stability and control},
  volume={10},
  number={1},
  pages={7--1},
  year={2007}
}

@book{sauer2017power,
  title={Power system dynamics and stability: with synchrophasor measurement and power system toolbox},
  author={Sauer, Peter W and Pai, Mangalore A and Chow, Joe H},
  year={2017},
  publisher={John Wiley \& Sons}
}

@article{zhang2020review,
  title={Review on set-theoretic methods for safety verification and control of power system},
  author={Zhang, Yichen and Li, Yan and Tomsovic, Kevin and Djouadi, Seddik M and Yue, Meng},
  journal={IET Energy Systems Integration},
  volume={2},
  number={3},
  pages={226--234},
  year={2020},
  publisher={Wiley Online Library}
}

@article{althoff2021set,
  title={Set propagation techniques for reachability analysis},
  author={Althoff, Matthias and Frehse, Goran and Girard, Antoine},
  journal={Annual Review of Control, Robotics, and Autonomous Systems},
  volume={4},
  number={1},
  pages={369--395},
  year={2021},
  publisher={Annual Reviews}
}

@inproceedings{chen2022reachability,
  title={Reachability analysis for cyber-physical systems: Are we there yet?},
  author={Chen, Xin and Sankaranarayanan, Sriram},
  booktitle={NASA formal methods symposium},
  pages={109--130},
  year={2022},
  organization={Springer}
}

@misc{lara2023powersimulationsdynamics,
      title={PowerSimulationsDynamics.jl -- An Open Source Modeling Package for Modern Power Systems with Inverter-Based Resources}, 
      author={Jose Daniel Lara and Rodrigo Henriquez-Auba and Matthew Bossart and Duncan S. Callaway and Clayton Barrows},
      year={2023},
      eprint={2308.02921},
      archivePrefix={arXiv},
      primaryClass={eess.SY},
      doi={https://doi.org/10.48550/arXiv.2308.02921}
}

@book{yazdani2010voltage,
  title={Voltage-sourced converters in power systems: modeling, control, and applications},
  author={Yazdani, Amirnaser and Iravani, Reza},
  year={2010},
  publisher={John Wiley \& Sons}
}

@article{hindmarsh2005sundials,
  title={SUNDIALS: Suite of nonlinear and differential/algebraic equation solvers},
  author={Hindmarsh, Alan C and Brown, Peter N and Grant, Keith E and Lee, Steven L and Serban, Radu and Shumaker, Dan E and Woodward, Carol S},
  journal={ACM Transactions on Mathematical Software (TOMS)},
  volume={31},
  number={3},
  pages={363--396},
  year={2005},
  publisher={ACM New York, NY, USA}
}

@article{lara2021powersystems,
  title={PowerSystems. jl—A power system data management package for large scale modeling},
  author={Lara, Jos{\'e} Daniel and Barrows, Clayton and Thom, Daniel and Krishnamurthy, Dheepak and Callaway, Duncan},
  journal={SoftwareX},
  volume={15},
  pages={100747},
  year={2021},
  publisher={Elsevier}
}

\appendices

\onecolumn

\section{IEEE 39-Bus Test System and Initialization}
\label{app:ieee39_init}
Both the analytical model and the $dq0$ EMT model are initialized from a common Newton--Raphson AC power-flow solution on the modified IEEE 39-bus system, computed with convergence tolerance $\epsilon_{\mathrm{PF}} = 10^{-10}$\,p.u.\ Network data are specified in MATPOWER format and solved by \textsc{PyPower} \texttt{runpf} for the analytical model and by \textsc{PowerFlows.jl} \texttt{ACPowerFlow} for the EMT model, with the converged bus-voltage magnitudes and phase angles defining the pre-disturbance operating point in both cases. For the analytical model, which adopts a DC network approximation, the AC solution provides the pre-disturbance active-power injections that define the equilibrium reference point from which all deviations are measured, and the bus-admittance matrix $Y_{\text{bus}} \in \mathbb{C}^{N \times N}$, assembled from the MATPOWER branch impedance and shunt data, from which the network susceptance Laplacian $\mathcal{B} = -\operatorname{Im}(Y_{\text{bus}}) \in \mathbb{R}^{N \times N}$ is extracted. The AC solution further confirms that pre-disturbance voltage magnitudes remain sufficiently close to $1.0$\,p.u. to support the DC approximation. Since the analytical model is formulated in deviation coordinates relative to this operating point, all dynamic states are initialized at zero. The EMT model, by contrast, is initialized directly from the full AC power-flow solution.

\subsection{Base Quantities}
\noindent\textbf{Network base:} The system base is $S_{\mathrm{sys}} = 100~\mathrm{MVA}$; all network quantities are expressed on this base.

\noindent\textbf{Base frequency:} The nominal frequency is $f_0 = 60~\mathrm{Hz}$, with base angular frequency $\Omega_b = 2\pi f_0$.

\noindent\textbf{Device base:} Each generator or inverter uses its own rating $S_g$ (MVA); machine internal parameters such as inertia, reactances, and time constants are specified on the corresponding device base.

\subsection{IEEE 39-Bus System Case Data}
\label{app:system_data}

This section reports the bus, generator, and branch case data for the modified IEEE 39-bus test system.

\subsubsection{Bus Data}
\begin{center}
\scriptsize
\setlength{\tabcolsep}{3.5pt}
\renewcommand{\arraystretch}{1.1}
\setlength{\LTleft}{0pt}
\setlength{\LTright}{0pt}

\begin{longtable}{@{\extracolsep{\fill}}@{}rrrrrrrrr@{}}
\caption{Bus data of the modified IEEE 39-bus case: bus type (1 = PQ, 2 = PV, 3 = Reference), loads in MW/MVAr, voltages in p.u., and nominal voltage levels in kV.}
\label{tab:bus_data}\\
\toprule
Bus & Type & $P_D$ & $Q_D$ & $V_m$ & $\theta$ (deg) & $V_{\max}$ & $V_{\min}$ & kV \\
\midrule
\endfirsthead
\toprule
Bus & Type & $P_D$ & $Q_D$ & $V_m$ & $\theta$ (deg) & $V_{\max}$ & $V_{\min}$ & kV \\
\midrule
\endhead
\bottomrule
\endlastfoot
1  & 1 & 97.6  & 44.2  & 1.047 & -8.439  & 1.06 & 0.94 & 345 \\
2  & 1 & 0.0   & 0.0   & 1.049 & -5.754  & 1.06 & 0.94 & 345 \\
3  & 1 & 322.0 & 2.4   & 1.030 & -8.599  & 1.06 & 0.94 & 345 \\
4  & 1 & 500.0 & 184.0 & 1.004 & -9.607  & 1.06 & 0.94 & 345 \\
5  & 1 & 0.0   & 0.0   & 1.005 & -8.612  & 1.06 & 0.94 & 345 \\
6  & 1 & 0.0   & 0.0   & 1.008 & -7.950  & 1.06 & 0.94 & 345 \\
7  & 1 & 233.8 & 84.0  & 0.997 & -10.124 & 1.06 & 0.94 & 345 \\
8  & 1 & 522.0 & 176.0 & 0.996 & -10.615 & 1.06 & 0.94 & 345 \\
9  & 1 & 6.5   & -66.6 & 1.028 & -10.322 & 1.06 & 0.94 & 345 \\
10 & 1 & 0.0   & 0.0   & 1.017 & -5.427  & 1.06 & 0.94 & 345 \\
11 & 1 & 0.0   & 0.0   & 1.013 & -6.284  & 1.06 & 0.94 & 345 \\
12 & 1 & 7.5   & 88.0  & 1.000 & -6.244  & 1.06 & 0.94 & 138 \\
13 & 1 & 0.0   & 0.0   & 1.014 & -6.098  & 1.06 & 0.94 & 345 \\
14 & 1 & 0.0   & 0.0   & 1.012 & -7.656  & 1.06 & 0.94 & 345 \\
15 & 1 & 320.0 & 153.0 & 1.015 & -7.736  & 1.06 & 0.94 & 345 \\
16 & 1 & 329.0 & 32.3  & 1.032 & -6.188  & 1.06 & 0.94 & 345 \\
17 & 1 & 0.0   & 0.0   & 1.034 & -7.301  & 1.06 & 0.94 & 345 \\
18 & 1 & 158.0 & 30.0  & 1.031 & -8.224  & 1.06 & 0.94 & 345 \\
19 & 1 & 0.0   & 0.0   & 1.050 & -1.023  & 1.06 & 0.94 & 345 \\
20 & 1 & 628.0 & 103.0 & 0.991 & -2.015  & 1.06 & 0.94 & 230 \\
21 & 1 & 274.0 & 115.0 & 1.032 & -3.781  & 1.06 & 0.94 & 345 \\
22 & 1 & 0.0   & 0.0   & 1.050 & 0.668   & 1.06 & 0.94 & 345 \\
23 & 1 & 247.5 & 84.6  & 1.045 & 0.470   & 1.06 & 0.94 & 345 \\
24 & 1 & 308.6 & -92.2 & 1.037 & -6.068  & 1.06 & 0.94 & 345 \\
25 & 1 & 224.0 & 47.2  & 1.058 & -4.363  & 1.06 & 0.94 & 345 \\
26 & 1 & 139.0 & 17.0  & 1.052 & -5.527  & 1.06 & 0.94 & 345 \\
27 & 1 & 281.0 & 75.5  & 1.038 & -7.495  & 1.06 & 0.94 & 345 \\
28 & 1 & 206.0 & 27.6  & 1.050 & -2.015  & 1.06 & 0.94 & 345 \\
29 & 1 & 283.5 & 26.9  & 1.050 & 0.744   & 1.06 & 0.94 & 345 \\
30 & 2 & 20.0  & 10.0  & 1.048 & -3.334  & 1.06 & 0.94 & 16.5 \\
31 & 2 & 9.2   & 4.6   & 0.982 & 0.000   & 1.06 & 0.94 & 16.5 \\
32 & 2 & 0.0   & 0.0   & 0.983 & 2.569   & 1.06 & 0.94 & 16.5 \\
33 & 2 & 0.0   & 0.0   & 0.997 & 4.195   & 1.06 & 0.94 & 16.5 \\
34 & 2 & 0.0   & 0.0   & 1.012 & 3.175   & 1.06 & 0.94 & 16.5 \\
35 & 2 & 20.0  & 10.0  & 1.049 & 5.630   & 1.06 & 0.94 & 16.5 \\
36 & 2 & 0.0   & 0.0   & 1.064 & 8.323   & 1.06 & 0.94 & 16.5 \\
37 & 2 & 0.0   & 0.0   & 1.028 & 2.421   & 1.06 & 0.94 & 16.5 \\
38 & 2 & 0.0   & 0.0   & 1.026 & 7.808   & 1.06 & 0.94 & 16.5 \\
39 & 3 & 0.0   & 0.0   & 1.030 & 0.000   & 1.06 & 0.94 & 345 \\
\end{longtable}
\end{center}

\subsubsection{Generator Data}
\begin{center}
\scriptsize
\setlength{\tabcolsep}{4pt}
\renewcommand{\arraystretch}{1.1}
\setlength{\LTleft}{0pt}
\setlength{\LTright}{0pt}

\begin{longtable}{@{\extracolsep{\fill}}@{}rrrrrrrrr@{}}
\caption{Generator case data of the modified IEEE 39-bus system: active/reactive injections in MW/MVAr, device base in MVA, and active/reactive operating limits.}
\label{tab:gen_data}\\
\toprule
Bus & $P_G$ & $Q_G$ & $Q_{\max}$ & $Q_{\min}$ & $V_G$ & $S_{\mathrm{base}}$ & $P_{\max}$ & $P_{\min}$ \\
\midrule
\endfirsthead
\toprule
Bus & $P_G$ & $Q_G$ & $Q_{\max}$ & $Q_{\min}$ & $V_G$ & $S_{\mathrm{base}}$ & $P_{\max}$ & $P_{\min}$ \\
\midrule
\endhead
\bottomrule
\endlastfoot
30 & 250.0  & 146.16 & 400 & 140  & 1.048 & 1000  & 840  & 0 \\
31 & 520.8  & 198.25 & 300 & -100 & 0.982 & 1000  & 646  & 0 \\
32 & 650.0  & 205.14 & 300 & 150  & 0.983 & 1000  & 725  & 0 \\
33 & 632.0  & 109.91 & 250 & 0    & 0.997 & 1000  & 652  & 0 \\
34 & 508.0  & 165.76 & 167 & 0    & 1.012 & 600   & 508  & 0 \\
35 & 650.0  & 212.41 & 300 & -100 & 1.049 & 1000  & 687  & 0 \\
36 & 560.0  & 101.18 & 240 & 0    & 1.064 & 1000  & 580  & 0 \\
37 & 540.0  & 0.44   & 250 & 0    & 1.028 & 1000  & 564  & 0 \\
38 & 830.0  & 22.84  & 300 & -150 & 1.026 & 1000  & 865  & 0 \\
39 & 1000.0 & 88.28  & 300 & -100 & 1.030 & 10000 & 1100 & 0 \\
\end{longtable}
\end{center}

\subsubsection{Branch Data}
\begin{center}
\scriptsize
\setlength{\tabcolsep}{3.8pt}
\renewcommand{\arraystretch}{1.1}
\setlength{\LTleft}{0pt}
\setlength{\LTright}{0pt}

\begin{longtable}{@{\extracolsep{\fill}}@{}rrrrrrrr@{}}
\caption{Branch case data of the modified IEEE 39-bus system: series resistance $R$, reactance $X$, total line charging susceptance $B$, thermal rating, tap ratio, and phase shift. All quantities are given on the 100 MVA system base. Entries with nonzero \textsc{tap} correspond to transformer branches.}
\label{tab:branch_data}\\
\toprule
From & To & $R$ & $X$ & $B$ & MVA & TAP & SHIFT \\
\midrule
\endfirsthead
\toprule
From & To & $R$ & $X$ & $B$ & MVA & TAP & SHIFT \\
\midrule
\endhead
\bottomrule
\endlastfoot
1  & 2  & 0.004 & 0.041 & 0.698 & 600  & 0     & 0 \\
1  & 39 & 0.001 & 0.025 & 0.750 & 1000 & 0     & 0 \\
2  & 3  & 0.001 & 0.015 & 0.258 & 500  & 0     & 0 \\
2  & 25 & 0.007 & 0.009 & 0.146 & 500  & 0     & 0 \\
3  & 4  & 0.001 & 0.021 & 0.222 & 500  & 0     & 0 \\
3  & 18 & 0.001 & 0.013 & 0.214 & 500  & 0     & 0 \\
4  & 5  & 0.001 & 0.013 & 0.134 & 600  & 0     & 0 \\
4  & 14 & 0.001 & 0.013 & 0.138 & 500  & 0     & 0 \\
5  & 6  & 0.000 & 0.003 & 0.044 & 1200 & 0     & 0 \\
5  & 8  & 0.001 & 0.011 & 0.148 & 900  & 0     & 0 \\
6  & 7  & 0.001 & 0.009 & 0.114 & 900  & 0     & 0 \\
6  & 11 & 0.001 & 0.008 & 0.138 & 480  & 0     & 0 \\
7  & 8  & 0.000 & 0.005 & 0.078 & 900  & 0     & 0 \\
8  & 9  & 0.002 & 0.036 & 0.380 & 900  & 0     & 0 \\
9  & 39 & 0.001 & 0.025 & 1.200 & 900  & 0     & 0 \\
10 & 11 & 0.000 & 0.004 & 0.072 & 600  & 0     & 0 \\
10 & 13 & 0.000 & 0.004 & 0.072 & 600  & 0     & 0 \\
13 & 14 & 0.001 & 0.010 & 0.172 & 600  & 0     & 0 \\
14 & 15 & 0.002 & 0.022 & 0.366 & 600  & 0     & 0 \\
15 & 16 & 0.001 & 0.009 & 0.172 & 600  & 0     & 0 \\
16 & 17 & 0.001 & 0.009 & 0.134 & 600  & 0     & 0 \\
16 & 19 & 0.002 & 0.020 & 0.304 & 600  & 0     & 0 \\
16 & 21 & 0.001 & 0.014 & 0.254 & 600  & 0     & 0 \\
16 & 24 & 0.000 & 0.006 & 0.068 & 600  & 0     & 0 \\
17 & 18 & 0.001 & 0.008 & 0.132 & 600  & 0     & 0 \\
17 & 27 & 0.001 & 0.017 & 0.322 & 600  & 0     & 0 \\
21 & 22 & 0.001 & 0.014 & 0.256 & 900  & 0     & 0 \\
22 & 23 & 0.001 & 0.010 & 0.184 & 600  & 0     & 0 \\
23 & 24 & 0.002 & 0.035 & 0.362 & 600  & 0     & 0 \\
25 & 26 & 0.003 & 0.032 & 0.512 & 600  & 0     & 0 \\
26 & 27 & 0.001 & 0.015 & 0.240 & 600  & 0     & 0 \\
26 & 28 & 0.004 & 0.047 & 0.780 & 600  & 0     & 0 \\
26 & 29 & 0.006 & 0.063 & 1.028 & 600  & 0     & 0 \\
28 & 29 & 0.001 & 0.015 & 0.250 & 600  & 0     & 0 \\
2  & 30 & 0.000 & 0.018 & 0.000 & 900  & 1.025 & 0 \\
6  & 31 & 0.000 & 0.025 & 0.000 & 900  & 1.070 & 0 \\
10 & 32 & 0.000 & 0.020 & 0.000 & 900  & 1.070 & 0 \\
12 & 11 & 0.002 & 0.043 & 0.000 & 500  & 1.006 & 0 \\
12 & 13 & 0.002 & 0.043 & 0.000 & 500  & 1.006 & 0 \\
19 & 20 & 0.001 & 0.014 & 0.000 & 900  & 1.060 & 0 \\
19 & 33 & 0.001 & 0.014 & 0.000 & 900  & 1.070 & 0 \\
20 & 34 & 0.001 & 0.018 & 0.000 & 900  & 1.009 & 0 \\
22 & 35 & 0.000 & 0.014 & 0.000 & 900  & 1.025 & 0 \\
23 & 36 & 0.000 & 0.027 & 0.000 & 900  & 1.000 & 0 \\
25 & 37 & 0.001 & 0.023 & 0.000 & 900  & 1.025 & 0 \\
29 & 38 & 0.001 & 0.016 & 0.000 & 900  & 1.025 & 0 \\
\end{longtable}
\end{center}

\section{Model Comparison of the Analytical and $dq0$-EMT Models}
\label{app:model_comparison}

\renewcommand{\arraystretch}{1.2}
\setlength{\tabcolsep}{3pt}
\footnotesize
\setlength{\LTleft}{0pt}
\setlength{\LTright}{0pt}

\begin{longtable}{@{}
>{\raggedright\arraybackslash}p{2.35cm}|
>{\raggedright\arraybackslash}p{6.75cm}|
>{\raggedright\arraybackslash}p{6.75cm}
@{}}
\caption{PSID $dq0$-EMT and analytical models across subsystems.}
\label{tab:model_comparison}\\
\toprule
\textbf{Component} & \textbf{Analytical model} & \textbf{$dq0$-EMT model (PSID.jl)} \\
\midrule
\endfirsthead

\toprule
\textbf{Component} & \textbf{Analytical model} & \textbf{$dq0$-EMT model (PSID.jl)} \\
\midrule
\endhead

\bottomrule
\endlastfoot

\multicolumn{3}{l}{\textit{\textbf{A. Synchronous Generators}}} \\
\midrule

\textbf{Rotor} &
$\dot{\theta}_g = \omega_g$
\newline
$M_g \dot{\omega}_g = -D_g \omega_g + p_{\mathrm{m},g} - p_{\ell,g} - p_g(\theta)$
\newline
$\omega_g$: frequency deviation from synchronous speed $\omega_s$; $p_{\ell,g}$: local load (zero if no load connected), $M_g$: SG inertia constant.
&
$\dot{\delta}_g = \Omega_b(\omega_g - \omega_s)$
\newline
$2H_g\,\dot{\omega}_g = \tau_{\mathrm{m},g} - \tau_{\mathrm{e},g} - D_g(\omega_g - \omega_s)$
\newline
$\omega_g$: per-unit rotor electrical speed, $\omega_s$: synchronous reference speed, $\Omega_b$: electrical base angular frequency.
\\

\midrule
\textbf{Flux} &
Internal electromagnetic states (e.g., $e'_q$, $e'_d$, $\psi_{kd}$, $\psi_{kq}$) evolve much faster than rotor-speed and governor dynamics; we therefore enforce quasi-steady subtransient behavior and retain only the electromechanical dynamics.
&
\texttt{RoundRotorQuadratic}
\newline
$\dot{e}'_q = \dfrac{1}{T'_{d0}}\!\left(v_f - X_{ad}I_{fd}\right)$,
\quad
$\dot{e}'_d = \dfrac{1}{T'_{q0}}\!\left(-X_{aq}I_{1q}\right)$
\newline
$\dot{\psi}_{kd} = \dfrac{1}{T''_{d0}}\!\left[-\psi_{kd} + e'_q - (x'_d - x_l)i_d\right]$
\newline
$\dot{\psi}_{kq} = \dfrac{1}{T''_{q0}}\!\left[-\psi_{kq} + e'_d + (x'_q - x_l)i_q\right]$
\\

\midrule
\textbf{AVR} &
AVRs are driven by electromagnetic dynamics, and voltage regulation evolves on a faster timescale than electromechanical frequency dynamics. Therefore, bus voltage magnitudes are fixed at their pre-disturbance operating-point values.
&
\texttt{AVRTypeI}
\newline
$\dot{v}_f = -\dfrac{1}{T_e}\!\left[v_f(K_e + S_e(v_f)) - v_{r1}\right]$
\newline
$\dot{v}_{r1} = \dfrac{1}{T_a}\!\left[K_a\!\left(v_{\mathrm{ref}} - v_m - v_{r2} - \dfrac{K_f}{T_f}v_f\right) - v_{r1}\right]$
\newline
$\dot{v}_{r2} = -\dfrac{1}{T_f}\!\left[\dfrac{K_f}{T_f}v_f + v_{r2}\right]$,
\quad
$\dot{v}_m = \dfrac{1}{T_r}(v_h - v_m)$
\newline
$S_e(v_f) = A_e \exp(B_e|v_f|)$
\\

\midrule
\textbf{Governor} &
\textbf{First-order lag:}
$\tau_g \dot{p}_{\mathrm{m},g} = -p_{\mathrm{m},g} + p^{r}_{\cG,g} - R_g^{-1}\omega_g$
&
\texttt{TGTypeI}
\newline
$\dot{x}_{g1} = \dfrac{1}{T_s}(p_{\mathrm{in}} - x_{g1})$
\newline
$\dot{x}_{g2} = \dfrac{1}{T_c}\!\left[\left(1 - \dfrac{T_3}{T_c}\right)x_{g1} - x_{g2}\right]$
\newline
$\dot{x}_{g3} = \dfrac{1}{T_5}\!\left[\left(1 - \dfrac{T_4}{T_5}\right)\!\left(x_{g2} + \dfrac{T_3}{T_c}x_{g1}\right) - x_{g3}\right]$
\newline
$p_{\mathrm{in}} = P_{\mathrm{ref}} + R^{-1}(\omega_{\mathrm{ref}} - \omega_g)$
\newline
$T_c = T_3 = T_4 = T_5 = \epsilon$ (small $\approx 0.001$), making dynamics effectively first-order via $T_s$, matching analytical $\tau_g$.
\\

\midrule
\textbf{PSS} & The damping contribution of the PSS is absorbed into $D_g$.
&
\texttt{IEEEST}
\newline
$A_4\dot{x}_1 = u - A_3x_1 - x_2$,
\quad
$\dot{x}_2 = x_1$
\newline
$A_2\dot{x}_3 = x_2 - A_1x_3 - x_4$,
\quad
$\dot{x}_4 = x_3$
\newline
$T_2\dot{x}_5 = \left(1 - \dfrac{T_1}{T_2}\right)y_f - x_5$
\newline
$T_4\dot{x}_6 = \left(1 - \dfrac{T_3}{T_4}\right)y_{LL1} - x_6$
\newline
$T_6\dot{x}_7 = -\left(K_s\dfrac{T_5}{T_6}y_{LL2} + x_7\right)$
\\

\midrule
\multicolumn{3}{l}{\textit{\textbf{B. Grid-Forming Inverters}}} \\
\midrule

\textbf{Outer control} &
$\dot{\theta}_d = \omega_d$
\newline
$M_d \dot{\omega}_d = -D_d \omega_d + p^{r}_{\cI,d} - p_{\ell,d} - p_d(\theta)$
\newline
$M_d = T_{f,d}/m_{p,d}$, \quad $D_d = 1/m_{p,d}$
\newline
$\theta_d$: GFM angle state, $\omega_d$: GFM frequency-deviation state, $M_d$: Virtual inertia constant, $D_d = 1/m_{p,d}$: Effective droop gain.
&
\texttt{VirtualInertia + ReactivePowerDroop}
\newline
$\dot{\omega}_{\mathrm{olc}} = \dfrac{1}{T_a}\!\left[p_{\mathrm{ref}} - p_e - k_d(\omega_{\mathrm{olc}} - \omega_{\mathrm{pll}}) - k_\omega(\omega_{\mathrm{olc}} - \omega_{\mathrm{ref}})\right]$
\newline
$\dot{\theta}_{\mathrm{olc}} = \Omega_b(\omega_{\mathrm{olc}} - \omega_{\mathrm{sys}})$
\newline
$\dot{q}_m = \omega_f(q_e - q_m)$
\newline
$p_e = v_ri_r + v_ii_i$,
\quad
$q_e = v_ii_r - v_ri_i$
\newline
$v^{\mathrm{ref}}_{\mathrm{olc}} = v^{\mathrm{ref}} + k_q(q^{\mathrm{ref}} - q_m)$
\newline
$T_a$: VSM inertia constant, $k_d$: VSM damping constant, $k_\omega$: frequency droop gain, $\omega_{\mathrm{olc}}$: outer-loop frequency state, $\omega_{\mathrm{ref}}$: nominal frequency reference. Since \texttt{FreqEstimator= None} for the GFM, $\omega_{\mathrm{pll}}=\omega_{\mathrm{fix}}=1.0$ p.u.
\\

\midrule
\textbf{Inner control} &
Inner voltage/current loops are assumed fast and quasi-steady.
&
\texttt{VoltageModeControl}
\newline
$\dot{\xi}_d = v^{\mathrm{ref}}_{d,\mathrm{vi}} - v_d$,
\quad
$\dot{\xi}_q = v^{\mathrm{ref}}_{q,\mathrm{vi}} - v_q$
\newline
$\dot{\gamma}_d = i^{\mathrm{ref}}_{d,\mathrm{cv}} - i_{d,\mathrm{cv}}$,
\quad
$\dot{\gamma}_q = i^{\mathrm{ref}}_{q,\mathrm{cv}} - i_{q,\mathrm{cv}}$
\newline
$\dot{\phi}_d = \omega_{\mathrm{ad}}(v_d - \phi_d)$,
\quad
$\dot{\phi}_q = \omega_{\mathrm{ad}}(v_q - \phi_q)$
\newline
$v^{\mathrm{ref}}_{d,\mathrm{vi}} = v^{\mathrm{ref}}_{\mathrm{olc}} - r_v i_d + \omega_{\mathrm{olc}} l_v i_q$
\newline
$v^{\mathrm{ref}}_{q,\mathrm{vi}} = -\,r_v i_q - \omega_{\mathrm{olc}} l_v i_d$
\newline
$i^{\mathrm{ref}}_{d,\mathrm{cv}} = k_{pv}(v^{\mathrm{ref}}_{d,\mathrm{vi}} - v_d) + k_{iv}\xi_d - c_f\omega_{\mathrm{olc}}v_q$
\newline
$i^{\mathrm{ref}}_{q,\mathrm{cv}} = k_{pv}(v^{\mathrm{ref}}_{q,\mathrm{vi}} - v_q) + k_{iv}\xi_q + c_f\omega_{\mathrm{olc}}v_d$
\newline
$v^{\mathrm{ref\text{-}sig}}_d = k_{pc}(i^{\mathrm{ref}}_{d,\mathrm{cv}} - i_{d,\mathrm{cv}}) + k_{ic}\gamma_d - \omega_{\mathrm{olc}} l_f i_{q,\mathrm{cv}} - k_{\mathrm{ad}}(v_d - \phi_d)$
\newline
$v^{\mathrm{ref\text{-}sig}}_q = k_{pc}(i^{\mathrm{ref}}_{q,\mathrm{cv}} - i_{q,\mathrm{cv}}) + k_{ic}\gamma_q + \omega_{\mathrm{olc}} l_f i_{d,\mathrm{cv}} - k_{\mathrm{ad}}(v_q - \phi_q)$
\\

\midrule
\textbf{LCL filter} &
Filter and interface dynamics are reduced by time-scale separation.
&
\texttt{LCLFilter}
\newline
$\dot{i}_{d,c} = \dfrac{\Omega_b}{\ell_f}\!\left(v_{\mathrm{cv},d} - v_d - r_f i_{d,c} + \omega_{\mathrm{sync}}\ell_f i_{q,c}\right)$
\newline
$\dot{v}_d = \dfrac{\Omega_b}{c_f}\!\left(i_{d,c} - i_d + \omega_{\mathrm{sync}}c_f v_q\right)$
\newline
$\dot{i}_d = \dfrac{\Omega_b}{\ell_g}\!\left(v_d - v_{\mathrm{grid},d} - r_g i_d + \omega_{\mathrm{sync}}\ell_g i_q\right)$
\newline
$\dot{i}_{q,c} = \dfrac{\Omega_b}{\ell_f}\!\left(v_{\mathrm{cv},q} - v_q - r_f i_{q,c} - \omega_{\mathrm{sync}}\ell_f i_{d,c}\right)$
\newline
$\dot{v}_q = \dfrac{\Omega_b}{c_f}\!\left(i_{q,c} - i_q - \omega_{\mathrm{sync}}c_f v_d\right)$
\newline
$\dot{i}_q = \dfrac{\Omega_b}{\ell_g}\!\left(v_q - v_{\mathrm{grid},q} - r_g i_q - \omega_{\mathrm{sync}}\ell_g i_d\right)$
\\

\midrule
\multicolumn{3}{l}{\textit{\textbf{C. Grid-Following Inverter}}} \\
\midrule

\textbf{PLL / frequency estimate} &
Frequency divider:
\[
\omega_{\mathcal F}=H_{\mathcal I}\omega_{\mathcal I}+H_{\mathcal G}\omega_{\mathcal G},
\]
where $\omega_{\mathcal F}$ is estimated algebraically from the swing-bus (GFM and SG) frequencies via the frequency-divider relation.
&
\texttt{KauraPLL(Frequency Estimator)}
\newline
$\dot{v}_{d,\mathrm{pll}} = \omega_{\mathrm{lp}}(v_{d,\mathrm{out}} - v_{d,\mathrm{pll}})$
\newline
$\dot{v}_{q,\mathrm{pll}} = \omega_{\mathrm{lp}}(v_{q,\mathrm{out}} - v_{q,\mathrm{pll}})$
\newline
$\dot{\varepsilon}_{\mathrm{pll}} = \tan^{-1}(v_{q,\mathrm{pll}}/v_{d,\mathrm{pll}})$
\newline
$\dot{\theta}_{\mathrm{pll}} = \Omega_b\,\delta\omega_{\mathrm{pll}}$
\newline
$\delta\omega_{\mathrm{pll}} = 1.0 - \omega_{\mathrm{sys}} + k_{p,\mathrm{pll}}\tan^{-1}(v_{q,\mathrm{pll}}/v_{d,\mathrm{pll}}) + k_{i,\mathrm{pll}}\varepsilon_{\mathrm{pll}}$
\\

\midrule
\textbf{Outer control} &
At bus $f\in\cF$, the GFL active-power injection is modeled as
\[
p_f \approx p^{r}_{\cF,f} - K_f \omega_f.
\]

Enforcing active-power balance at the bus gives
\[
p_f = p_{\ell,f} + p_f(\theta),
\]
and therefore
\[
0 = p^{r}_{\cF,f} - K_f \omega_f - p_{\ell,f} - p_f(\theta).
\]
where $K_f$ is the Fast Frequency Response (FFR) gain.
&
\texttt{ActivePowerPI + ReactivePowerPI}
\newline
$\omega_{\mathrm{olc}} = \omega_{\mathrm{pll}}, \qquad \theta_{\mathrm{olc}} = \theta_{\mathrm{pll}}$
\newline
$p = P_{\mathrm{ref}} - K_f(\omega_{\mathrm{pll}} - \omega_{\mathrm{ref}})$
\newline
$\dot{\sigma}_p = p - p_m$,
\quad
$\dot{p}_m = \omega_z(p_e - p_m)$
\newline
$\dot{\sigma}_q = q_{\mathrm{ref}} - q_m$,
\quad
$\dot{q}_m = \omega_f(q_e - q_m)$
\newline
$p_e = v_r i_r + v_i i_i$,
\quad
$q_e = v_i i_r - v_r i_i$
\newline
$i^{\mathrm{ref}}_{d,\mathrm{cv}} = k_{qp}(q_{\mathrm{ref}} - q_m) + k_{qi}\sigma_q$
\newline
$i^{\mathrm{ref}}_{q,\mathrm{cv}} = k_{pp}(p - p_m) + k_{pi}\sigma_p$
\\

\midrule
\textbf{Inner control} &
Current-control dynamics are assumed fast and quasi-steady.
&
\texttt{CurrentModeControl}
\newline
$\dot{\gamma}_d = i^{\mathrm{ref}}_{d,\mathrm{cv}} - i_{d,\mathrm{cv}}$
\newline
$\dot{\gamma}_q = i^{\mathrm{ref}}_{q,\mathrm{cv}} - i_{q,\mathrm{cv}}$
\newline
$v^{\mathrm{ref\text{-}signal}}_d = k_{pc}(i^{\mathrm{ref}}_{d,\mathrm{cv}} - i_{d,\mathrm{cv}}) + k_{ic}\gamma_d - \omega_{\mathrm{pll}} l_f i_{q,\mathrm{cv}} + k_{\mathrm{ffv}}v_d$
\newline
$v^{\mathrm{ref\text{-}signal}}_q = k_{pc}(i^{\mathrm{ref}}_{q,\mathrm{cv}} - i_{q,\mathrm{cv}}) + k_{ic}\gamma_q + \omega_{\mathrm{pll}} l_f i_{d,\mathrm{cv}} + k_{\mathrm{ffv}}v_q$
\\

\midrule
\textbf{LCL filter} &
Filter and interface dynamics are reduced by time-scale separation.
&
\texttt{LCLFilter}
\newline
Same filter structure as for GFM units, with $\omega_{\mathrm{sync}} = \omega_{\mathrm{pll}}$.
\\

\midrule
\multicolumn{3}{l}{\textit{\textbf{D. Transmission Lines}}} \\
\midrule

\textbf{TL models} &
\textbf{AC model:}
\[
I_i = \sum_{j=1}^{N} Y_{ij}V_j,
\qquad
Y_{ij} = G_{ij} + jB_{ij}
\]
\[
S_i = P_i + jQ_i = V_i I_i^{*},
\qquad
V_i = |V_i|e^{j\theta_i}
\]
\[
P_i = \sum_{j=1}^{N}|V_i||V_j|
\Big(G_{ij}\cos(\theta_i-\theta_j)+B_{ij}\sin(\theta_i-\theta_j)\Big)
\]
\[
Q_i = \sum_{j=1}^{N}|V_i||V_j|
\Big(G_{ij}\sin(\theta_i-\theta_j)-B_{ij}\cos(\theta_i-\theta_j)\Big)
\]
\textbf{DC approximation:}
$|V_i|\approx 1$ p.u., $G_{ij}\approx 0$,
$\cos(\theta_i-\theta_j)\approx 1$,
$\sin(\theta_i-\theta_j)\approx \theta_i-\theta_j$
\newline
\textbf{Line flow:} $p_{ij}(\theta)\approx b_{ij}(\theta_i-\theta_j)$
\[
p(\theta)=\mathcal{B}\theta_{\mathrm{bus}}
\]
&
\texttt{DynamicBranch} $\pi$-model in $dq0$
\newline
\[
\frac{1}{\Omega_b}
\begin{bmatrix}\mathbf{L}&\mathbf{0}\\ \mathbf{0}&\mathbf{L}\end{bmatrix}
\frac{d}{dt}
\begin{bmatrix}\vec{i}^{\,\ell}_d\\ \vec{i}^{\,\ell}_q\end{bmatrix}
=
\mathbf{E}_{\ell}
\begin{bmatrix}\vec{v}_d\\ \vec{v}_q\end{bmatrix}
-
\begin{bmatrix}\mathbf{R}&-\mathbf{L}\\ \mathbf{L}&\mathbf{R}\end{bmatrix}
\begin{bmatrix}\vec{i}^{\,\ell}_d\\ \vec{i}^{\,\ell}_q\end{bmatrix}
\]
\[
\frac{1}{\Omega_b}
\begin{bmatrix}\mathbf{B}^{-1}&\mathbf{0}\\ \mathbf{0}&\mathbf{B}^{-1}\end{bmatrix}
\frac{d}{dt}
\begin{bmatrix}\vec{v}^{\,b}_d\\ \vec{v}^{\,b}_q\end{bmatrix}
=
\begin{bmatrix}\vec{i}^{\,b}_d\\ \vec{i}^{\,b}_q\end{bmatrix}
-
\begin{bmatrix}\Re(\mathbf{Y}_a)&\Im(\mathbf{Y}_a)\\ -\Im(\mathbf{Y}_a)&\Re(\mathbf{Y}_a)\end{bmatrix}
\begin{bmatrix}\vec{v}^{\,b}_d\\ \vec{v}^{\,b}_q\end{bmatrix}
\]
\\

\midrule
\multicolumn{3}{l}{\textit{\textbf{E. Load Modeling}}} \\
\midrule

\textbf{Load model:} &
\textbf{Constant power load:}
\[
S_{d,k}=P_{d,k}+jQ_{d,k}, \qquad P_{d,k},Q_{d,k}>0.
\]
For a pure load bus, $P_{g,k}=Q_{g,k}=0$, so that
\[
P_k^{\mathrm{inj}}=-P_{d,k}, \qquad Q_k^{\mathrm{inj}}=-Q_{d,k}.
\]
Accordingly, the AC power-flow equations impose
\[
P_k^{\mathrm{inj}}(V,\theta)=-P_{d,k}, \qquad
Q_k^{\mathrm{inj}}(V,\theta)=-Q_{d,k}.
\]
Under the DC approximation, this reduces to the active-power balance
\[
0=-p_{\ell,n_\ell}-p_{n_\ell}(\theta), \qquad n_\ell\in\mathcal L.
\]
&
ZIP model with constant-power component
\newline
\[
S_{d,k}=P_{d,k}+jQ_{d,k}=v_k\,i_{\mathrm{load},k}^{*}
\]
\[
i_{\mathrm{load},k}=\frac{P_{d,k}-jQ_{d,k}}{v_k^{*}}
\]
Writing $v_k=v_{d,k}+jv_{q,k}$ and $i_{\mathrm{load},k}=i_{d,\mathrm{load},k}+j\,i_{q,\mathrm{load},k}$,
\[
i_{d,\mathrm{load},k}
=
\frac{P_{d,k}v_{d,k}+Q_{d,k}v_{q,k}}{v_{d,k}^{2}+v_{q,k}^{2}}
\]
\[
i_{q,\mathrm{load},k}
=
\frac{P_{d,k}v_{q,k}-Q_{d,k}v_{d,k}}{v_{d,k}^{2}+v_{q,k}^{2}}
\]
\\

\midrule
\end{longtable}

\section{Device Parameters for Time-Domain Simulation}
\label{app:device_params}

\noindent
\begin{minipage}{\linewidth}
\small
\textit{\textbf{Note. Bold entries denote parameters used in both the PSID $dq0$-EMT and analytical models}}.
\end{minipage}

\subsection{Synchronous-Machine Parameters (SG Buses 33, 34, 36, 37, 39)}

\vspace{4pt}
\footnotesize
\renewcommand{\arraystretch}{1.15}
\setlength{\tabcolsep}{3pt}
\setlength{\LTleft}{0pt}
\setlength{\LTright}{0pt}

\begin{longtable}{@{\extracolsep{\fill}}@{}cccccccccccccc@{}}
\caption{Synchronous-machine parameters (RoundRotorQuadratic model + SingleMass shaft).}
\label{tab:sg_machine}\\
\toprule
Bus & $R_s$ & $x_l$ & $x_d$ & $x_q$ & $x'_d$ & $x'_q$ & $x''_d=x''_q$ & $T'_{d0}$ (s) & $T'_{q0}$  & $T''_{d0}$ (s) & $T''_{q0}$  & $\mathbf{H}$ & $\mathbf{D}$  \\
\midrule
\endfirsthead

\toprule
Bus & $R_s$ & $x_l$ & $x_d$ & $x_q$ & $x'_d$ & $x'_q$ & $x''_d=x''_q$ & $T'_{d0}$ (s) & $T'_{q0}$  & $T''_{d0}$ (s) & $T''_{q0}$  & $\mathbf{H}$ & $\mathbf{D}$  \\
\midrule
\endhead

\bottomrule
\endlastfoot

33 & 0.02 & 0.236 & 2.096 & 2.064 & 0.349 & 1.328 & 0.280 & 5.690 & 1.500 & 0.050 & 0.035 & \textbf{2.860} & \textbf{40} \\
34 & 0.02 & 0.300 & 2.000 & 1.900 & 0.600 & 0.800 & 0.400 & 7.000 & 0.700 & 0.050 & 0.035 & \textbf{4.333} & \textbf{40} \\
36 & 0.02 & 0.225 & 2.065 & 2.044 & 0.343 & 1.302 & 0.308 & 5.660 & 1.500 & 0.050 & 0.035 & \textbf{3.480} & \textbf{40} \\
37 & 0.02 & 0.300 & 2.000 & 1.900 & 0.600 & 0.800 & 0.400 & 7.000 & 0.700 & 0.050 & 0.035 & \textbf{2.430} & \textbf{40} \\
39 & 0.02 & 0.300 & 2.000 & 1.900 & 0.600 & 0.800 & 0.400 & 7.000 & 0.700 & 0.050 & 0.035 & \textbf{5.000} & \textbf{48} \\
\end{longtable}

\subsection{SG Governor Parameters (TGTypeI)}
\noindent\textit{IEEE Type-I turbine governor. Parameters are uniform across all SG buses (33, 34, 36, 37, 39).}

\vspace{4pt}
\footnotesize
\renewcommand{\arraystretch}{1.15}
\setlength{\tabcolsep}{5pt}
\setlength{\LTleft}{0pt}
\setlength{\LTright}{0pt}

\begin{longtable}{@{\extracolsep{\fill}}@{}>{\raggedright\arraybackslash}p{5.6cm} c c @{}}
\caption{SG governor parameters (TGTypeI).}
\label{tab:sg_governor}\\
\toprule
Parameter & Symbol & Value \\
\midrule
\endfirsthead
\toprule
Parameter & Symbol & Value \\
\midrule
\endhead
\bottomrule
\endlastfoot
Speed droop & $\mathbf{R}$ & \textbf{0.045 pu} \\
Servo time constant & $\mathbf{T_s\,(\tau_g)}$ & \textbf{2.0 s}\\
Controller time constant & $T_c$ & 0.01 s \\
Lead--lag time constant 1 & $T_3$ & 0.01 s \\
Lead--lag time constant 2 & $T_4$ & 0.01 s \\
Lead--lag time constant 3 & $T_5$ & 0.01 s \\
Valve position limits & $[v_{\min},\,v_{\max}]$ & $[-5,\,+5]$ pu \\
\end{longtable}

\subsection{SG Exciter Parameters (AVRTypeI)}
\noindent\textit{IEEE Type-I automatic voltage regulator with exponential saturation.}

\vspace{4pt}
\footnotesize
\renewcommand{\arraystretch}{1.15}
\setlength{\tabcolsep}{3.5pt}
\setlength{\LTleft}{0pt}
\setlength{\LTright}{0pt}

\begin{longtable}{@{\extracolsep{\fill}}@{}cccccccccccc@{}}
\caption{SG exciter parameters (AVRTypeI).}
\label{tab:sg_avr}\\
\toprule
Bus & $K_a$ & $K_e$ & $K_f$ & $T_a$ (s) & $T_e$ (s) & $T_f$ (s) & $T_r$ (s) & $V_{a,\min}$ (pu) & $V_{a,\max}$ (pu) & $A_e$ & $B_e$ \\
\midrule
\endfirsthead
\toprule
Bus & $K_a$ & $K_e$ & $K_f$ & $T_a$ (s) & $T_e$ (s) & $T_f$ (s) & $T_r$ (s) & $V_{a,\min}$ (pu) & $V_{a,\max}$ (pu) & $A_e$ & $B_e$ \\
\midrule
\endhead
\bottomrule
\endlastfoot
33 & 1.0 & $-0.052$ & 0.0 & 0.015 & 0.500 & 1.0 & 0.010 & $-10$ & 10 & 0.001 & 1.0 \\
34 & 1.0 & 1.000 & 0.0 & 0.015 & 1.000 & 1.0 & 0.010 & $-10$ & 10 & 0.001 & 1.0 \\
36 & 1.0 & 1.000 & 0.0 & 0.015 & 0.730 & 1.0 & 0.010 & $-10$ & 10 & 0.001 & 1.0 \\
37 & 1.0 & 1.000 & 0.0 & 0.015 & 1.000 & 1.0 & 0.010 & $-10$ & 10 & 0.001 & 1.0 \\
39 & 1.0 & 1.000 & 0.0 & 0.015 & 1.000 & 1.0 & 0.010 & $-10$ & 10 & 0.001 & 1.0 \\
\end{longtable}

\subsection{SG Power-System Stabilizer Parameters (IEEEST)}
\noindent\textit{IEEE-type stabilizer applied uniformly to all SG buses.}

\vspace{4pt}
\footnotesize
\renewcommand{\arraystretch}{1.15}
\setlength{\tabcolsep}{6pt}
\setlength{\LTleft}{0pt}
\setlength{\LTright}{0pt}

\begin{longtable}{@{\extracolsep{\fill}}@{}>{\raggedright\arraybackslash}p{6.6cm} c c@{}}
\caption{SG power-system stabilizer (PSS) parameters (\texttt{IEEEST}).}
\label{tab:sg_pss}\\
\toprule
Parameter & Symbol & Value \\
\midrule
\endfirsthead
\toprule
Parameter & Symbol & Value \\
\midrule
\endhead
\bottomrule
\endlastfoot
Input signal code & --- & 1 (rotor speed $\omega$) \\
Stabilizer gain & $K_s$ & 10.0 \\
Lead--lag 1 numerator & $T_1$ & 0.25 s \\
Lead--lag 1 denominator & $T_2$ & 0.02 s \\
Lead--lag 2 numerator & $T_3$ & 0.25 s \\
Lead--lag 2 denominator & $T_4$ & 0.02 s \\
Output-block time constants & $T_5,\;T_6$ & $1.0~\text{s},\;0.08~\text{s}$ \\
PSS output limits & $[L_{s,\min},\,L_{s,\max}]$ & $[-0.05,\,+0.05]$ pu \\
Upper cutoff bound & $V_{cu}$ & 1.25 pu \\
Lower cutoff bound & $V_{cl}$ & 1.00 pu \\
\end{longtable}

\subsection{Grid-Forming Inverter Parameters (GFM Buses 31, 32, 35, 38)}
\noindent\textit{Outer control: VirtualInertia + ReactivePowerDroop. Inner control: VoltageModeControl. Filter: LCLFilter.}

\vspace{4pt}
\footnotesize
\renewcommand{\arraystretch}{1.15}
\setlength{\tabcolsep}{3pt}
\setlength{\LTleft}{0pt}
\setlength{\LTright}{0pt}

\begin{longtable}{@{\extracolsep{\fill}}@{}ccccccccc@{}}
\caption{GFM inverter virtual-inertia, outer-control, and inner-control parameters (buses 31, 32, 35, and 38).}
\label{tab:gfm_params}\\
\toprule
Bus & $\mathbf{T_a\,(M_d)}$ (s) & $\mathbf{R_{\mathrm{GFM}}\,(m_{p,d})}$ (pu) & $\mathbf{k_\omega\,(D_d)}$ (pu) & $k_d$ & $k_{pv}$ & $k_{iv}$ & $k_{pc}$ & $k_{ic}$ \\
\midrule
\endfirsthead

\toprule
Bus & $\mathbf{T_a\,(M_d)}$ (s) & $\mathbf{R_{\mathrm{GFM}}\,(m_{p,d})}$ (pu) & $\mathbf{k_\omega\,(D_d)}$ (pu) & $k_d$ & $k_{pv}$ & $k_{iv}$ & $k_{pc}$ & $k_{ic}$ \\
\midrule
\endhead

\bottomrule
\endlastfoot

31 & \textbf{3.030} & \textbf{0.055} & \textbf{18.18 (5.5\% droop)} & 20 & 1.0 & 1000 & 3.0 & 600 \\
32 & \textbf{3.580} & \textbf{0.055} & \textbf{18.18 (5.5\% droop)} & 20 & 1.0 & 1000 & 3.0 & 600 \\
35 & \textbf{3.480} & \textbf{0.055} & \textbf{18.18 (5.5\% droop)} & 20 & 1.0 & 1000 & 3.0 & 600 \\
38 & \textbf{3.450} & \textbf{0.055} & \textbf{18.18 (5.5\% droop)} & 20 & 1.0 & 1000 & 3.0 & 600 \\
\end{longtable}

\vspace{4pt}
\footnotesize
\renewcommand{\arraystretch}{1.15}
\setlength{\tabcolsep}{6pt}
\setlength{\LTleft}{0pt}
\setlength{\LTright}{0pt}

\begin{longtable}{@{\extracolsep{\fill}}@{}>{\raggedright\arraybackslash}p{7.2cm} c c@{}}
\caption{LCL-filter, damping, and reactive-power-droop parameters for all GFM buses.}
\label{tab:gfm_filter}\\
\toprule
Parameter & Symbol & Value \\
\midrule
\endfirsthead
\toprule
Parameter & Symbol & Value \\
\midrule
\endhead
\bottomrule
\endlastfoot

Filter inductance (converter side) & $\ell_f$ & 0.08 pu \\
Filter resistance & $r_f$ & 0.003 pu \\
Filter capacitance & $c_f$ & 0.074 pu \\
Grid-side inductance & $\ell_g$ & 0.20 pu \\
Grid-side resistance & $r_g$ & 0.01 pu \\
Reactive-power droop gain & $k_q$ & 0.20 \\
Reactive-power filter bandwidth & $\omega_f$ & 1000 rad/s \\
Active damping cutoff frequency & $\omega_{\mathrm{ad}}$ & 100 rad/s \\
Active damping gain & $k_{\mathrm{ad}}$ & 2.0 \\
DC source voltage & $V_{\mathrm{DC}}$ & 1200 V \\
\end{longtable}

\subsection{Grid-Following Inverter Parameters (GFL Bus 30)}
\noindent\textit{Outer control: ActivePowerPI + ReactivePowerPI. Inner control: CurrentModeControl. Frequency estimator: KauraPLL. Filter: LCLFilter.}

\vspace{4pt}
\footnotesize
\renewcommand{\arraystretch}{1.15}
\setlength{\tabcolsep}{6pt}
\setlength{\LTleft}{0pt}
\setlength{\LTright}{0pt}

\begin{longtable}{@{\extracolsep{\fill}}@{}>{\raggedright\arraybackslash}p{7.4cm} c c@{}}
\caption{GFL inverter parameters (bus 30).}
\label{tab:gfl_params}\\
\toprule
Parameter & Symbol & Value \\
\midrule
\endfirsthead
\toprule
Parameter & Symbol & Value \\
\midrule
\endhead
\bottomrule
\endlastfoot

\textit{PLL (KauraPLL)} & & \\
Low-pass filter bandwidth & $\omega_{\mathrm{lp}}$ & 500 rad/s \\
Proportional gain & $k_{p,\mathrm{pll}}$ & 0.084 \\
Integral gain & $k_{i,\mathrm{pll}}$ & 4.69 \\

\midrule
\textit{Outer Active-Power Control (ActivePowerPI)} & & \\
\textbf{Fast Frequency-response (FFR) gain } & $\mathbf{K_f}$ & \textbf{0.055 pu} \\
Reference frequency & $\omega_{\mathrm{ref}}$ & 1.0 pu \\
Proportional gain & $k_{pp}$ & 2.0 \\
Integral gain & $k_{pi}$ & 30.0 \\
Zero frequency & $\omega_z$ & 500 rad/s \\

\midrule
\textit{Outer Reactive-Power Control (ReactivePowerPI)} & & \\
Proportional gain & $k_{qp}$ & 2.0 \\
Integral gain & $k_{qi}$ & 30.0 \\
Filter bandwidth & $\omega_f$ & 500 rad/s \\

\midrule
\textit{Inner Current Control (CurrentModeControl)} & & \\
Proportional gain & $k_{pc}$ & 3.0 \\
Integral gain & $k_{ic}$ & 600 \\
Feed-forward voltage gain & $k_{\mathrm{ffv}}$ & 0.0 \\

\midrule
\textit{LCL Filter} & & \\
Filter inductance / resistance / capacitance & $\ell_f,\,r_f,\,c_f$ & 0.08 / 0.003 / 0.074 pu \\
Grid inductance / resistance & $\ell_g,\,r_g$ & 0.20 / 0.01 pu \\
DC source voltage & $V_{\mathrm{DC}}$ & 1200 V \\
\end{longtable}

\noindent\small\textit{In the analytical model, the GFL frequency at bus 30 is determined algebraically via the frequency-divider relation}
\[
\omega_{30}=H_{30,\cI}\omega_{\cI}+H_{30,\cG}\omega_{\cG},
\]
\textit{and the active-power injection is represented by}
\[
0 = p^r_{\cF,30} - K_f \omega_{30} - p_{\ell,30} - p_{30}(\theta).
\]

\section{Additional Simulations for Scenarios 1 and 2 in Section~\ref{sec:simulations}}
\label{sec:additional_simulations}

In this appendix, we report additional simulation results for Scenarios~1 and~2 in Section~\ref{sec:simulations}. The figures show, for representative buses, the reachable tubes together with dq0-EMT and analytical-model trajectories.

\begin{figure*}[t]
    \centering
    \includegraphics[width=1.0\textwidth,keepaspectratio]{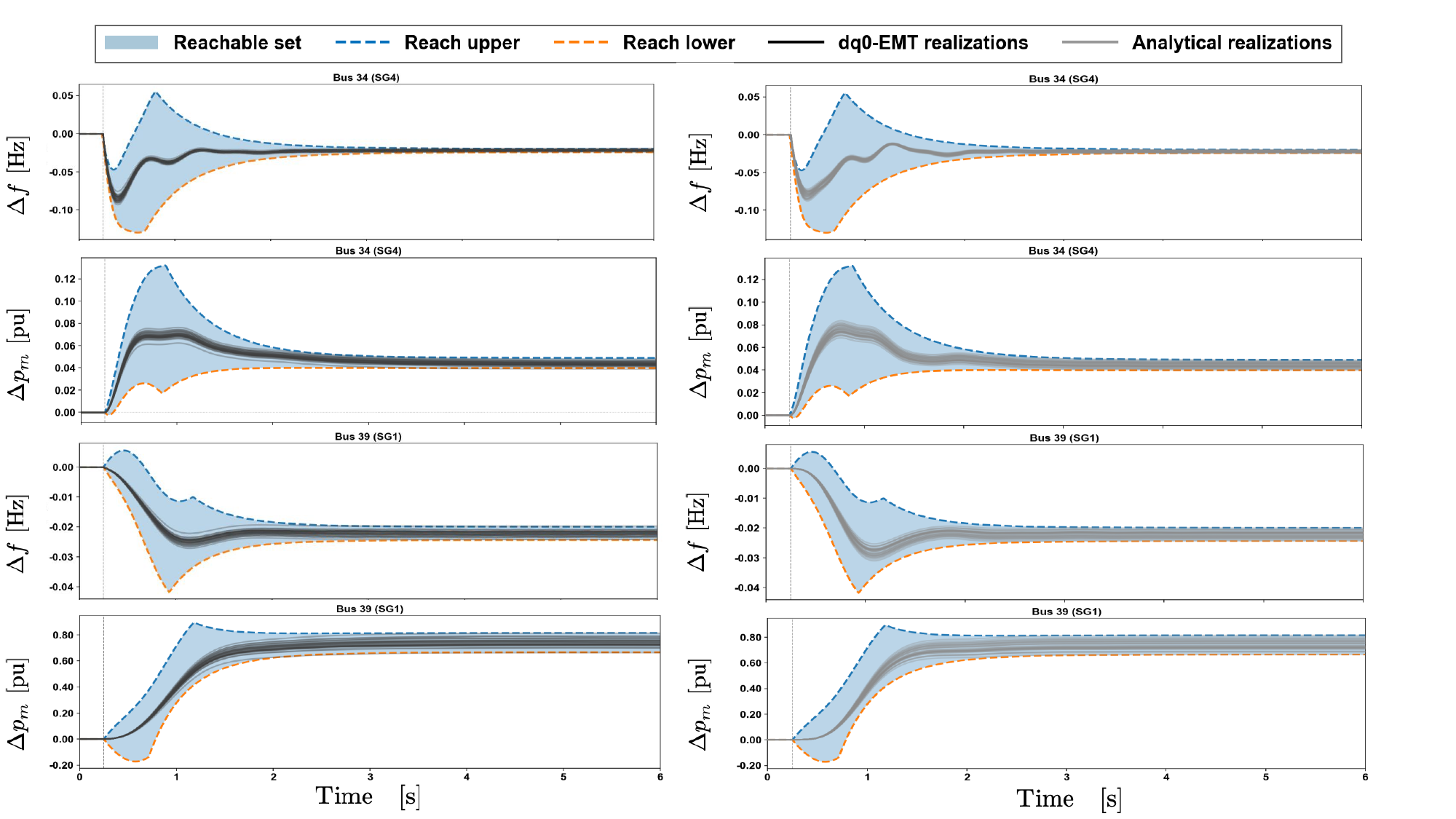}
    \caption{Reachable sets and trajectories for Scenario~1 at representative SG buses. The left panel shows dq0-EMT realizations, and the right panel shows analytical-model realizations. Vertical dashed lines indicate the load-change events.}
    \label{fig:scenario1SG}
    \vspace{-.4cm}
\end{figure*}

\begin{figure*}[t]
    \centering
    \includegraphics[width=1.0\textwidth,keepaspectratio]{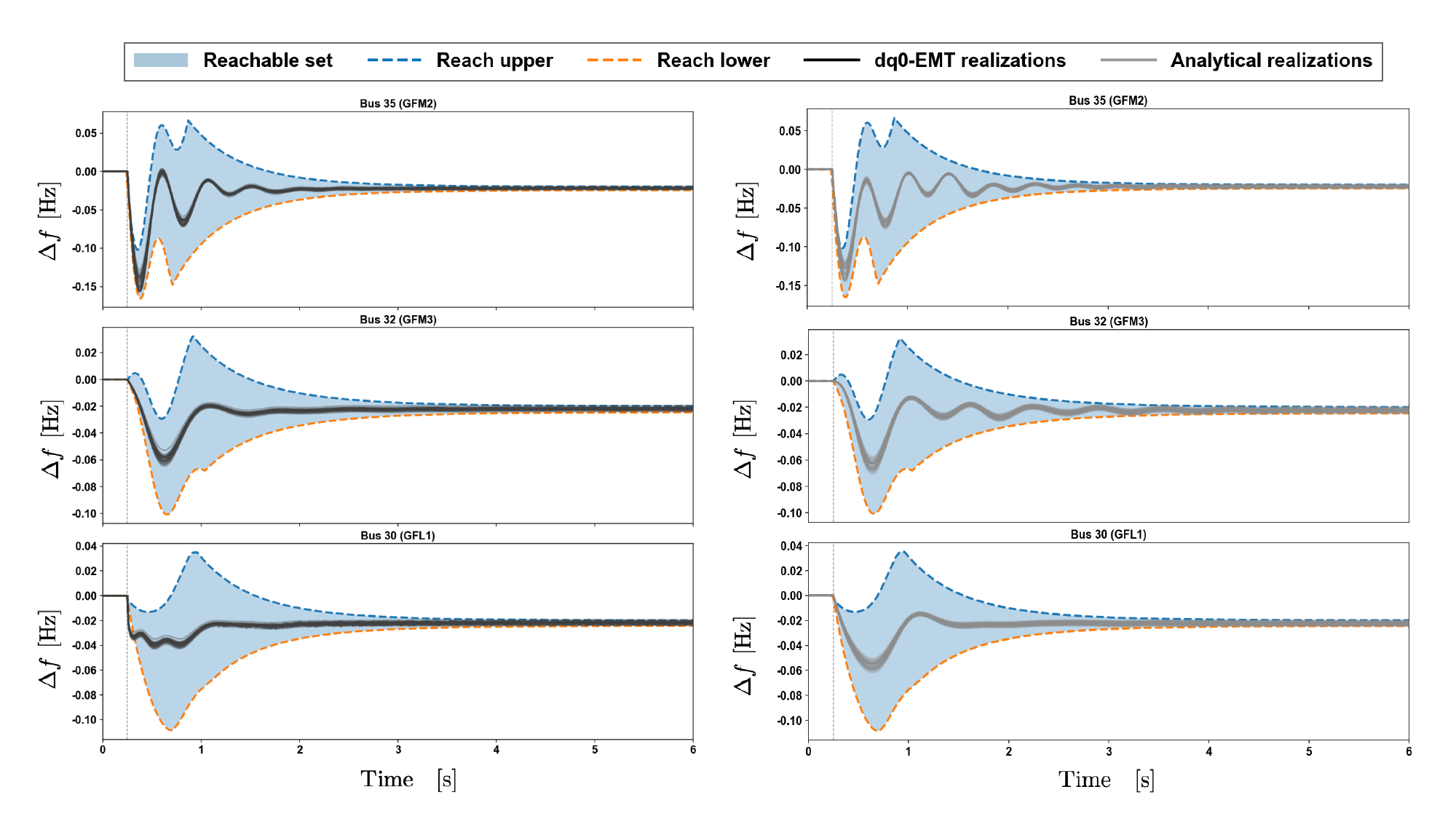}
    \caption{Reachable sets and trajectories for Scenario~1 at representative GFM and GFL buses. The left panel shows dq0-EMT realizations, and the right panel shows analytical-model realizations. Vertical dashed lines indicate the load-change events.}
    \label{fig:scenario1IBR}
    \vspace{-.4cm}
\end{figure*}

\begin{figure*}[t]
    \centering
    \includegraphics[width=1.0\textwidth,keepaspectratio]{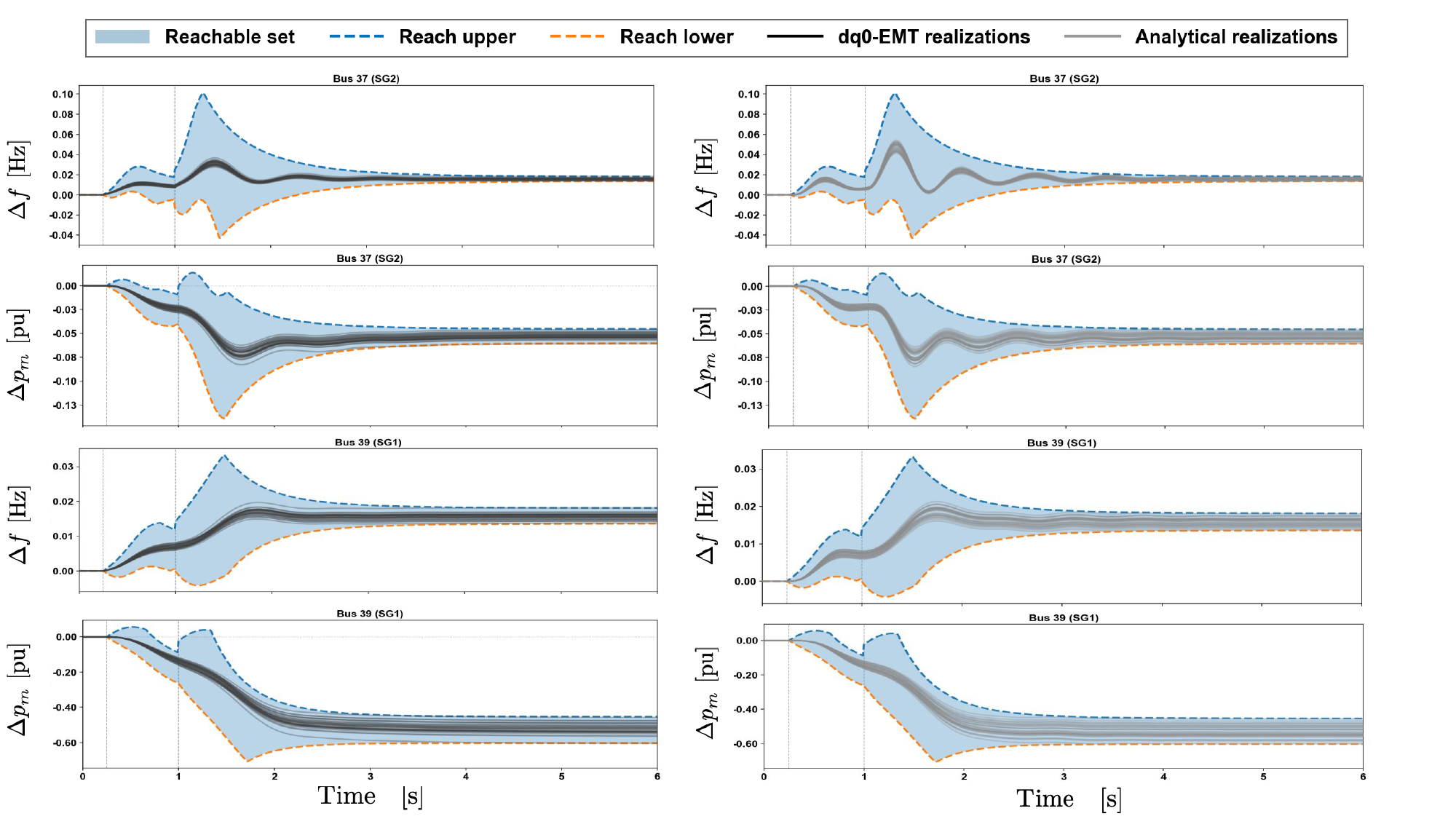}
    \caption{Reachable sets and trajectories for Scenario~2 at representative SG buses. The left panel shows dq0-EMT realizations, and the right panel shows analytical-model realizations. Vertical dashed lines indicate the load-change events.}
    \label{fig:scenario2SG}
    \vspace{-.4cm}
\end{figure*}

\begin{figure*}[t]
    \centering
    \includegraphics[width=1.0\textwidth,keepaspectratio]{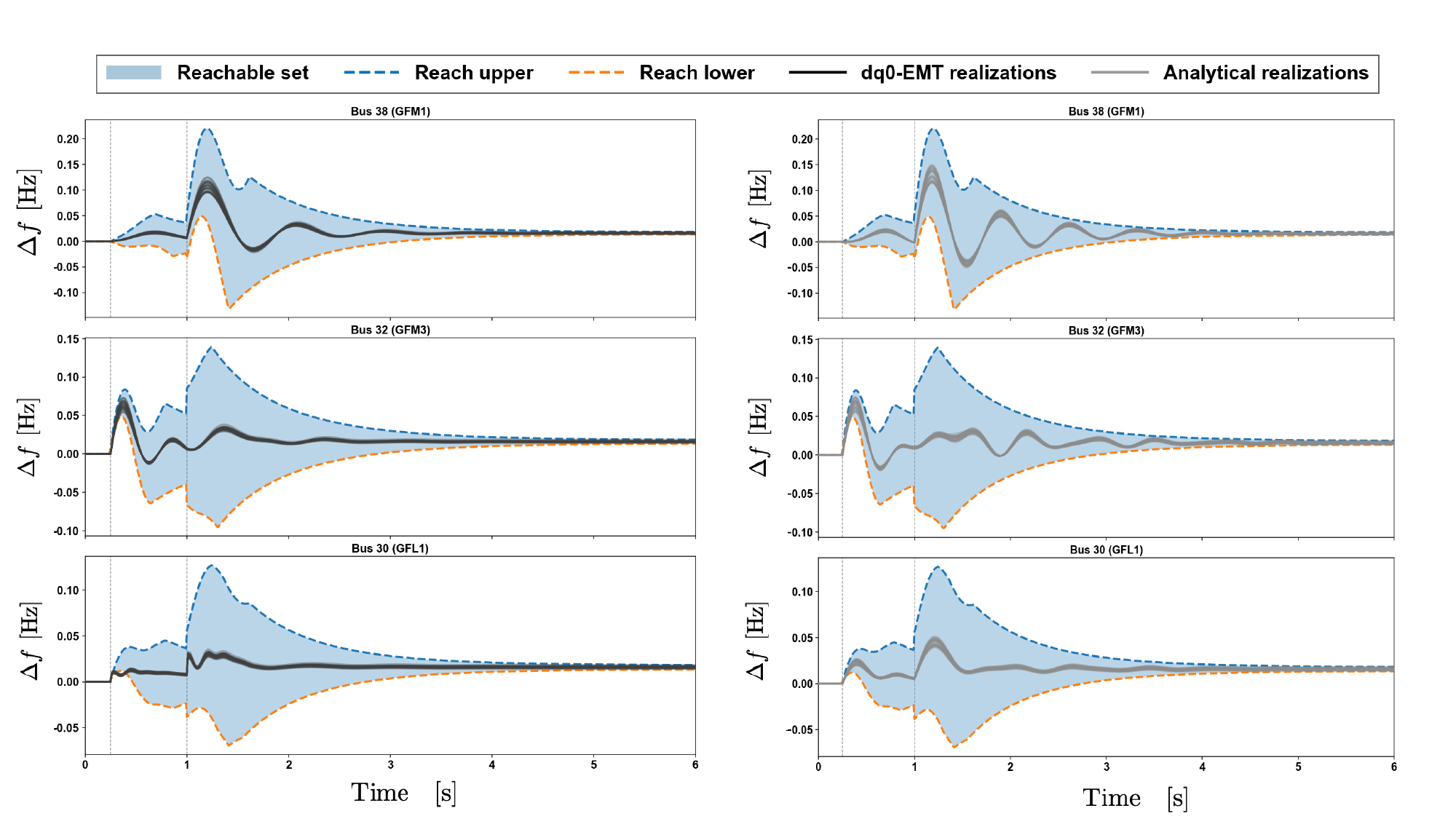}
    \caption{Reachable sets and trajectories for Scenario~2 at representative GFM and GFL buses. The left panel shows dq0-EMT realizations, and the right panel shows analytical-model realizations. Vertical dashed lines indicate the load-change events.}
    \label{fig:scenario2IBR}
    \vspace{-.4cm}
\end{figure*}

\end{document}